\documentclass[12pt,a4paper]{article}

\usepackage{epsfig}
\usepackage{amsmath,amsfonts,amssymb}
\usepackage{cite}
\usepackage{colortbl}
\usepackage{pifont}

\providecommand{\openone}{\leavevmode\hbox{\small1\kern-3.8pt\normalsize1}}

\newcommand{\ntrk}{N_\text{trk}}

\begin{document}

\begin{center}
\begin{Large}
{\bf Traces of a triboson resonance}
\end{Large}

\vspace{0.5cm}
J.~A.~Aguilar--Saavedra$^{a}$, J. H. Collins$^{b,c,d}$, S. Lombardo$^{d}$ \\[1mm]
\begin{small}
{$^a$ Departamento de F\'{\i}sica Te\'orica y del Cosmos, 
Universidad de Granada, \\ E-18071 Granada, Spain} \\ 
{$^b$ Maryland Center for Fundamental Physics, University of Maryland, College Park, MD 20742} \\
{$^c$ Department of Physics and Astronomy, Johns Hopkins University, Baltimore, \\ MD 21218, USA} \\
{$^d$  Department of Physics, LEPP, Cornell University, Ithaca, NY 14853, USA}
\end{small}
\end{center}

\begin{abstract}
We show that the relatively small but coincident excesses observed around 2 TeV in the ATLAS Run 1 and Run 2 hadronic diboson searches --- when a cut on the number of tracks in the fat jets is not applied --- and the null results of all remaining high-mass diboson searches are compatible with the decay of a triboson resonance $R$ into $WZ$ plus an extra particle $X$. These decays can take place via new neutral ($Y^0$) or charged ($Y^\pm$) particles, namely $R \to Y^0 \, W$, with $Y^0 \to Z X$, or $R \to Y^\pm Z$, with $Y^\pm \to W X$. An obvious candidate for such intermediate particle is a neutral one $Y^0$, given a $3.9\sigma$ excess found at 650 GeV by the CMS Collaboration in searches for intermediate mass diboson resonances decaying to $ZV$, with $V=W,Z$. We discuss discovery strategies for triboson resonances with small modifications of existing hadronic searches.

\end{abstract}

\section{Introduction}

New physics may show up in the searches performed at the Large Hadron Collider (LHC) in unexpected ways. The ATLAS and CMS experiments perform a large number of measurements in a variety of final states targetting simple ``benchmark'' novel signatures, but, if a sign of new physics beyond the Standard Model (SM) eventually appears, its real source may well be different from the specific signature searched for. In the LHC Run 1 at a centre-of-mass (CM) energy of 8 TeV the most remarkable deviation from the SM prediction was a $3.4\sigma$ excess found by the ATLAS Collaboration~\cite{Aad:2015owa} in the search for heavy resonances decaying into two SM gauge bosons $V=W/Z$ --- also known as diboson resonances ---  with the bosons decaying hadronically, giving rise to two fat jets $J$.  The excess was compatible with a new heavy resonance with a mass around 2 TeV, and it was largest in the $WZ$ channel where the two bosons were respectively tagged as a $W$ and a $Z$ boson. An analogous search performed by the CMS Collaboration~\cite{Khachatryan:2014hpa} also found an excess at the $2\sigma$ level around this mass. But searches performed in the semileptonic decay modes of the $WZ$ diboson pair\cite{Aad:2015ufa,Aad:2014xka,Khachatryan:2014gha}, some of them more sensitive than the hadronic mode, showed no hint of an excess at this mass.

The absence of any other signals motivated the proposal~\cite{Aguilar-Saavedra:2015rna} that the anomaly might be due to a triboson resonance, namely, a heavy resonance $R$ with a mass around 2 TeV decaying into one gauge boson plus an intermediate neutral or charged particle, $R \to Y^0 \, W$ or $R \to Y^\pm Z$, or perhaps both, where the masses of the intermediate particles $Y^0$, $Y^\pm$ could in principle lie in a wide range $300-1000$ GeV. The subsequent decays of $Y^0$ and $Y^\pm$ into a gauge boson plus some extra particle $X$, $Y^0 \to Z X$ and $Y^\pm \to W X$, would produce a $WZX$ triboson signal. And, for this signal, the presence of the extra particle $X$, which could be the SM Higgs boson $H$, a $W/Z$ boson or a new particle with a mass $M_X \lesssim 300$ GeV, would make the searches in the semileptonic channels much less sensitive than for a diboson resonance $R \to WZ$. Such a $WZX$ signal can be realised, for example, in left-right (LR) models when the ratio of the vacuum expectation values (VEVs) of the two scalars $\phi_1$, $\phi_2$ in the bidoublet, $\tan \beta \equiv v_2/v_1$, is either very small or very large~\cite{Aguilar-Saavedra:2015iew}, that is, one of the VEVs is much larger than the other. In that case, the $W'$ boson plays the role of the heavy resonance $R$ and the decays $W' \to WZ$, with a rate proportional to $\sin^2 2\beta$, are absent. But the decays into heavy scalars, e.g. $W' \to WH_1^0$, $W' \to H^\pm Z$, with a rate proportional to $\cos^2 2\beta$, are allowed. The subsequent cascade decay of the heavy scalars $H_1^0 \to Z A^0$, $H^\pm \to W A^0$, with $A^0$ a lighter pseudo-scalar, or $H_1^0 \to Z H$, $H^\pm \to W H$, yield triboson signals.

At the LHC Run 2, with a CM energy of 13 TeV, searches for diboson resonances have been performed using data taken in 2015. Searches in the semileptonic channels~\cite{ATLASlvJ2015,ATLASllJ2015,ATLASvvJ2015,CMS:2015nmz} have not shown any deviation from the SM prediction at this mass and searches in the hadronic channel~\cite{CMS:2015nmz,ATLASJJ2015} have not confirmed the previously seen excess.
Still, the data are not conclusive enough, and the examination of experimental results in the context of triboson signals is worthwhile. This is the main purpose of this paper. In section~\ref{sec:2} we discuss in detail the measurements performed by the ATLAS Collaboration in their searches for diboson resonances decaying hadronically, with their similarities and differences. In section~\ref{sec:3} we present our detailed Monte Carlo calculations of the QCD dijet background, which do not show any trace of a bump caused by the jet tagging criteria. Section~\ref{sec:4} is devoted to presenting predictions for several selected benchmark scenarios of triboson resonances in a variety of diboson searches. In this respect, the appearance of a $3.9\sigma$ excess in a CMS search for low-mass diboson resonances~\cite{CMS:2016tio} decaying into two opposite-sign leptons and a boson-tagged jet ($\ell \ell J$), at a mass of 650 GeV, suggests that the cascade decays of the heavy resonance can be mediated by a neutral particle $Y^0$ with this mass. In section~\ref{sec:5} we propose discovery strategies for triboson resonances in the fully hadronic ATLAS search, with minimal changes with respect to the analysis currently carried out. In section~\ref{sec:6} we discuss our results and their implications for the heavy neutral gauge boson that should accompany the charged resonance presumably responsible for a $WZX$ signal. The details of our simulations are given in appendix~\ref{sec:recast}, and a study of the impact on triboson signals of an upper cut on the number of jet tracks is presented in appendix~\ref{sec:a}. An addendum is included in appendix~\ref{sec:add} with the predictions of our triboson scenarios for the $VH$ hadronic search, whose experimental results were released by the ATLAS Collaboration after the submission of this paper.

\section{Closer look at fat dijet measurements}
\label{sec:2}

\begin{figure}[htb]
\begin{center}
\begin{tabular}{ccc}
\includegraphics[height=5.2cm,clip=]{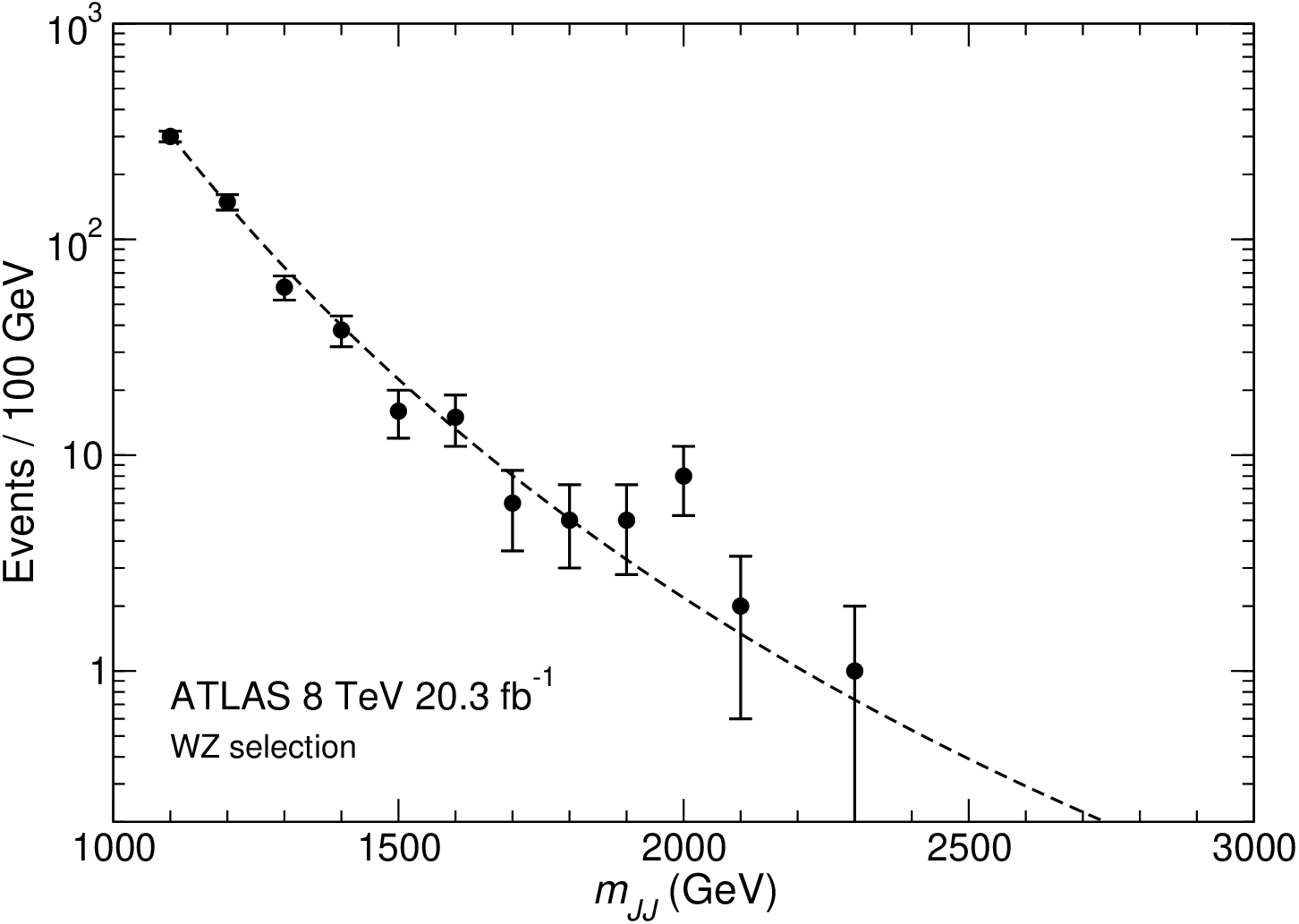} & \quad &
\includegraphics[height=5.2cm,clip=]{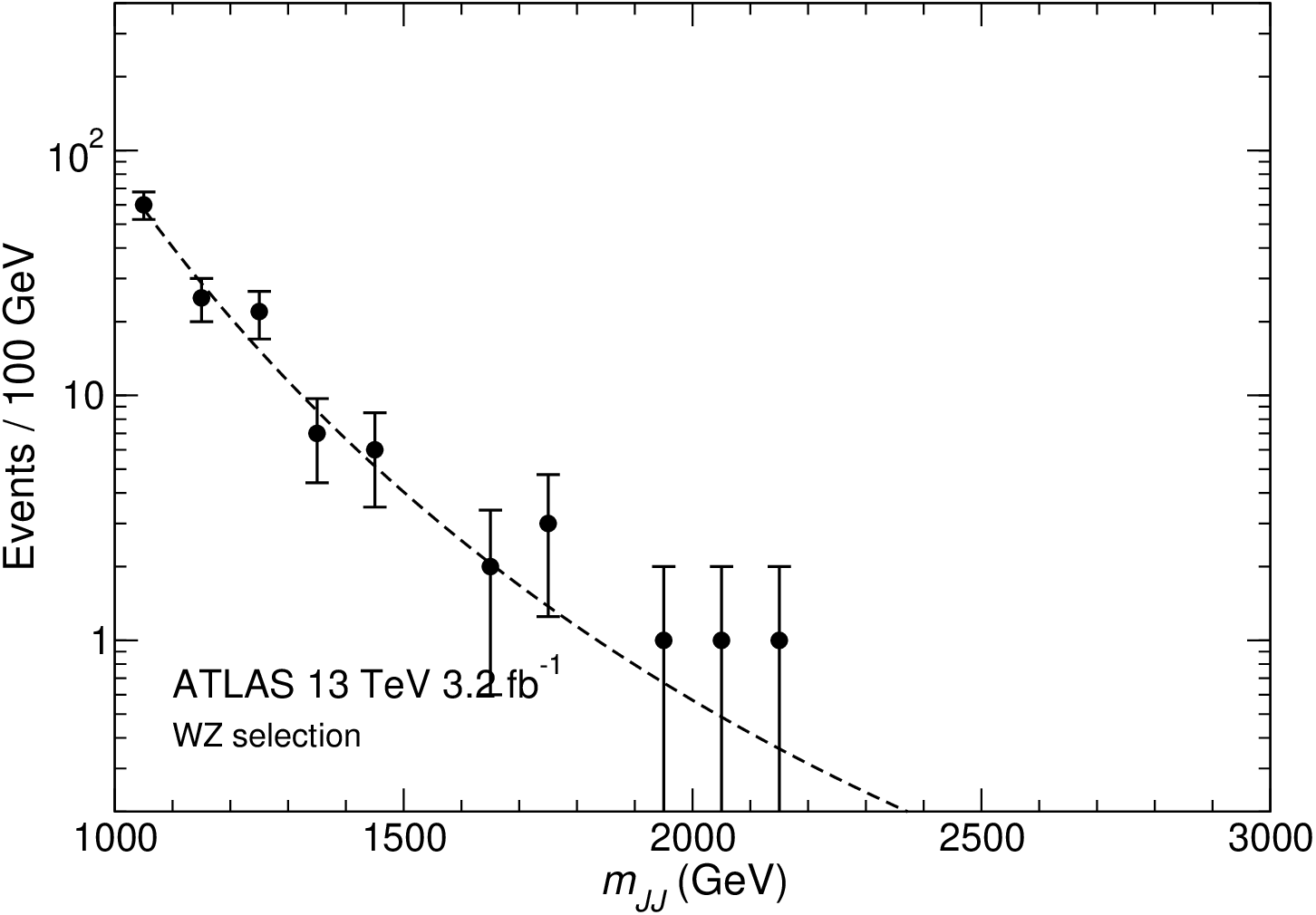}
\end{tabular}
\caption{Number of events with the nominal ATLAS $WZ$ selection in Run 1 (left) and Run 2 (right). The dashed lines correspond to the background-only fit.}
\label{fig:ATLASfs}
\end{center}
\end{figure}

As aforementioned, the ATLAS Collaboration found a $3.4\sigma$ excess in Run 1 data at 8 TeV with a luminosity of 20.3 fb$^{-1}$. For the reader's convenience, we reproduce in figure~\ref{fig:ATLASfs} (left) the number of events with the nominal $WZ$ ATLAS selection, with the background-only best fit. With Run 2 data at 13 TeV, the ATLAS Collaboration performed a search using 3.2 fb$^{-1}$. Given the similar efficiencies of both searches and the $7-8$ times larger $q \bar q$ partonic luminosity at 13 TeV, which compensates the smaller luminosity, a similar excess of around 8 events at 2 TeV was expected in Run 2, but the data, reproduced in figure~\ref{fig:ATLASfs} (right), show no significant deviation and are compatible with the background-only hypothesis at the $1\sigma$ level.

\begin{figure}[htb]
\begin{center}
\begin{tabular}{ccc}
\includegraphics[height=5.2cm,clip=]{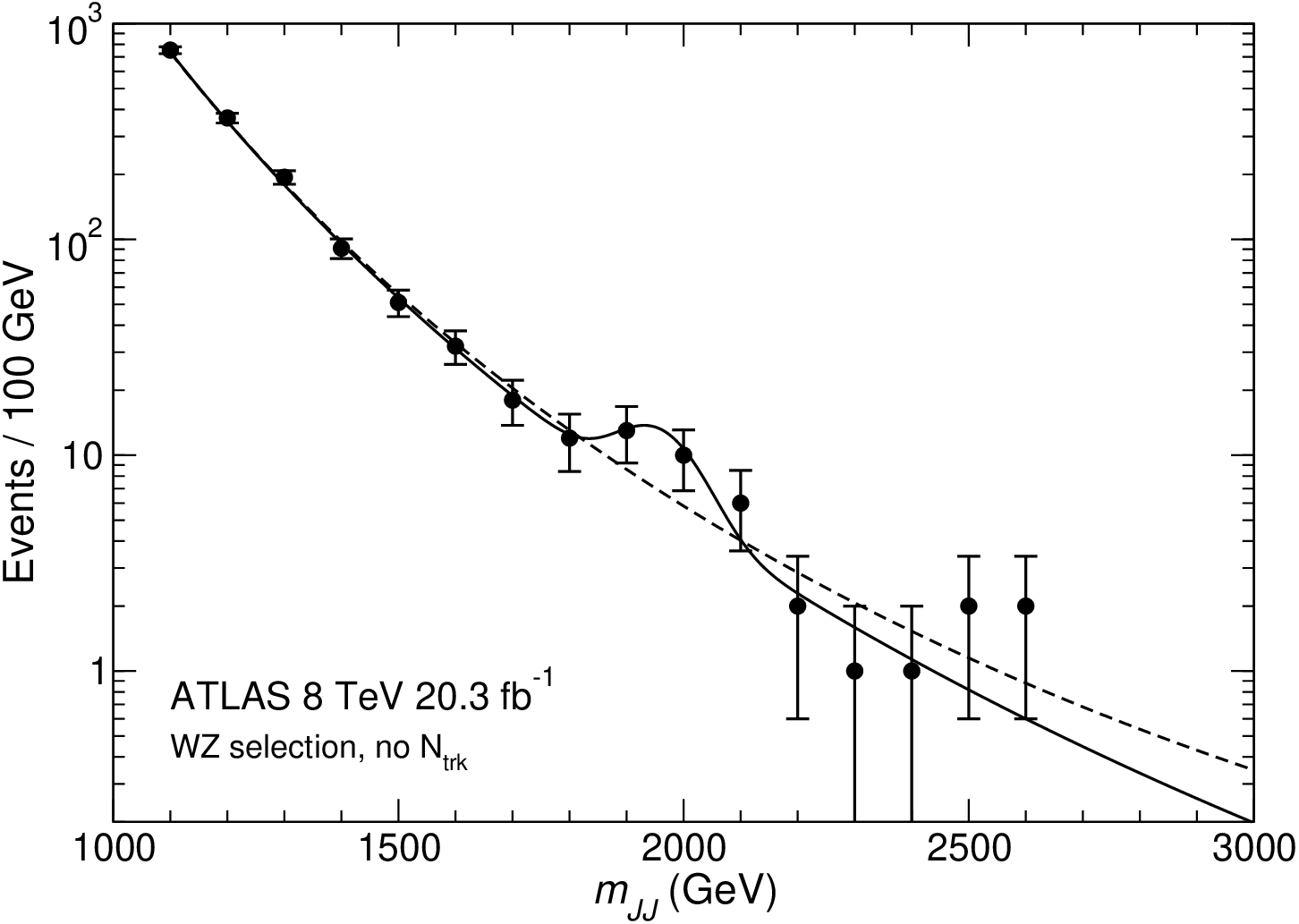} & \quad &
\includegraphics[height=5.2cm,clip=]{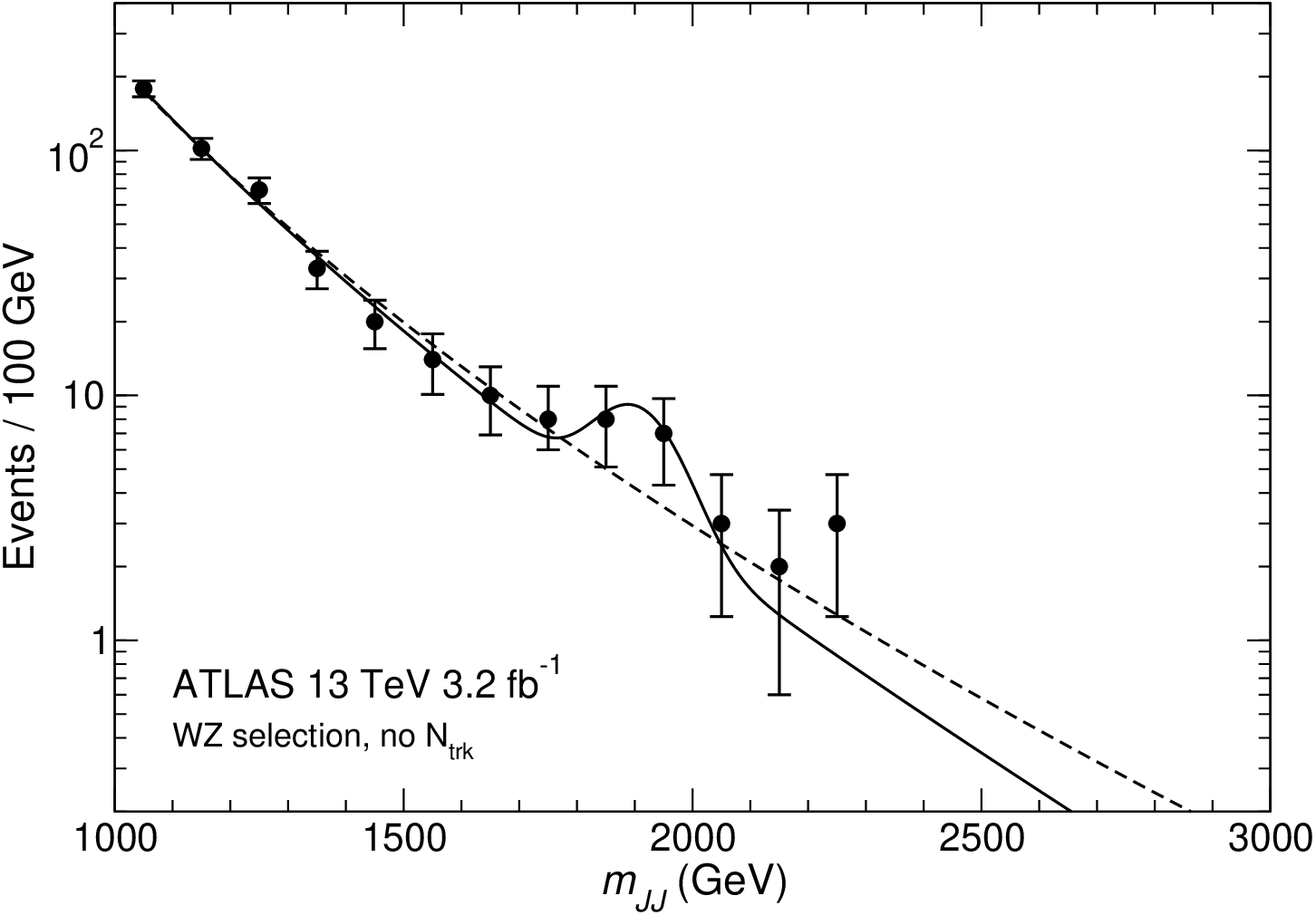}
\end{tabular}
\caption{Number of events with the nominal ATLAS $WZ$ selection except the $\ntrk$ cut in Run 1 (left) and Run 2 (right). The dashed lines correspond to the background-only fit. The solid lines show a simple signal plus background fit (see the text).}
\label{fig:ATLASnoNtrk}
\end{center}
\end{figure}

Besides data with the nominal selection, the ATLAS Collaboration has released results when one of the $V$ boson jet tagging requirements is dropped. We recall here that in the Run 1 analysis fat jets are reconstructed with the Cambridge-Aachen algorithm~\cite{Dokshitzer:1997in} with radius $R=1.2$, while in the Run 2 analysis the jets are reconstructed with the anti-$k_T$ algorithm~\cite{Cacciari:2008gp} with $R=1.0$. Several requirements are made on the reconstructed fat jets to be tagged as $W$ or $Z$ jets (for further details see refs.~\cite{Aad:2015owa,ATLASJJ2015}).
\begin{itemize}
\item A cut $\sqrt y \geq 0.45$ on the $y$ variable~\cite{Butterworth:2008iy} measuring the subjet momentum balance, which is replaced by a cut on the so-called $D_2$ function~\cite{Larkoski:2014gra} in Run 2. The precise value of the cut on $D_2$ depends on the jet transverse momentum.
\item A cut on the jet mass: $|m_J - M_V| < 13$ GeV, with $M_V = 82.4$ GeV (92.8 GeV) for $W$ ($Z$) bosons in Run 1. The cut is $|m_J - M_V| < 15$ GeV, with $M_V = 83.2$ GeV (93.4 GeV) in Run 2.
\item A requirement on the number of tracks with transverse momentum $p_T > 0.5$ GeV and originating from the primary vertex, $\ntrk < 30$.
\end{itemize}
Interestingly enough, Run 2 data when one of these three jet tagging requirements is removed also display bumps around 2 TeV, as it has already been pointed out~\cite{Aguilar-Saavedra:2015iew}. We will focus on results without the $\ntrk$ cut, as the corresponding numbers of data events are the closest to the ones with the nominal selection (in other words, removal of this requirement yields a smaller decrease of the expected signal significance than either the removal of the jet mass or the removal of the $\sqrt y / D_2$ cuts). We give in figure~\ref{fig:ATLASnoNtrk} the number of events in the Run 1~\cite{ATLASrun1ext} and Run 2~\cite{ATLASJJ2015} analyses for the $WZ$ selection without $\ntrk$. 
The dashed lines are the background-only best-fit, calculated with a maximum likelihood fit using the same background parameterisation of the ATLAS Collaboration. For illustration we also include simple signal plus background fits using for the signal a Gaussian with centre 1950 GeV (1900 GeV) for Run 1 (Run 2) and a width of 70 GeV in both cases. The estimated statistical significance of the bumps in the data, evaluated from the likelihood, is $2.5\sigma$ at Run 1 and $2.4\sigma$ at Run 2. The approximate size of the excesses found from the signal plus background fit to data is quite compatible: 13 events in Run 1 and 10 events in Run 2.

At this point, the urgent question arises of why, if the data with the $WZ$ selection without the $\ntrk$ cut show a consistent (but still not statistically significant) excess in both analyses, the results with the nominal selection (including the $\ntrk$ cut) are somewhat different. Let us discuss different possibilities in turn.

{\it Statistics only.} The coincidence of the the bumps at 8 and 13 TeV, plus some other excesses around 2 TeV like the CMS excess in $eejj$ production~\cite{Khachatryan:2014dka}, disfavour the hypothesis that all these bumps are statistical fluctuations in data. Still, this is an open possibility given the relatively small significance of the bumps. In any case, new data will elucidate whether these excesses are merely statistical artifacts or not.

{\it Mismodeling effects.} It has been pointed out~\cite{Martin:2016jdw} that the jet tagging requirements may cause a slope change in the continuum QCD dijet distribution. However, as it can be seen by eye in figure~\ref{fig:ATLASnoNtrk}, the observed bumps do not seem to be a slope change. As we will see in section~\ref{sec:3}, there is a slight change of slope in the Monte Carlo prediction for the QCD background. This change may affect the apparent size of a possible signal but does not give rise to a bump in the distributions. Besides, if the excess events at the bumps correspond to the very same QCD background as at the side bands, it is hard to conceive why in Run 1 data the further application of the $\ntrk$ cut shapes a peak, whereas in Run 2 data the $\ntrk$ cut flattens the bump.

{\it New physics.} The efficiencies of the $\ntrk$ cut have been evaluated by the ATLAS Collaboration using diboson signals and will generally be different for a different signal, {\it e.g.} a triboson. In order to explain the cleaning up of the bump in Run 2 data without resorting to large statistical fluctuations, a new physics signal would be required for which (i) the combination of cuts on jet mass, $D_2$ and $\ntrk$ for $R=1.0$ jets in Run 2 severely decreases the efficiency; (ii) but the combination of cuts on jet mass, $\sqrt y$ and $\ntrk$ for $R=1.2$ jets in Run 1 does not. In this regard, we have explored in appendix~\ref{sec:a} several $WZX$ triboson signals in several phase space regions and have found that, after the application of the remaining boson tagging cuts, the effect of the $\ntrk$ cut is very similar for the Run 1 and Run 2 analyses. 

{\it New physics and statistical fluctuations.} On the other hand, given the small statistics of the samples with the nominal $WZ$ selection, it might be possible that there is a downward fluctuation in Run 2 data and perhaps an upward fluctuation un Run 1 data. This possibility will, again, be tested when more data are available.

{\it New physics and mismodeling effects.} It is also possible that for new physics different from diboson resonances, the jet tagging variables are not correctly modeled by Monte Carlo simulations. Here it is worth pointing out that $\ntrk$ is not well modeled by simulation~\cite{Aad:2014gea}. In the Run 1 analysis the ATLAS Collaboration derives a scale factor of $0.9$ to correct the efficiency of the $\ntrk$ cut for $W$ and $Z$ bosons given by the simulation, while in the Run 2 analysis the number of tracks given by the simulation is multiplied by $1.07$. Despite these corrections being of the order of 10\%, it is conceivable that for fat jets from the multiboson cascade decay of a TeV-scale particle the agreement between simulation and data is worse, maybe with some unknown correlation between $\ntrk$ and other tagging variables. We also point out that the CMS Collaboration has chosen not to use $\ntrk$ to improve the expected signal significance, neither in the Run 1 nor in the Run 2 searches.

From the above discussion, we can conclude that the most likely explanations of the found excesses, if they persist with more data and become statistically significant, are (i) a new physics signal, for example a triboson, but also with some Monte Carlo mismodeling effect in jet substructure variables, perhaps in $\ntrk$; (ii) some new physics signal far more complex than a triboson, in which case one would also have to justify why the excesses seen in the ATLAS searches are localised.

\section{The QCD dijet background}
\label{sec:3}

In order to obtain a prediction for the QCD background, we fully recast Run 1 and Run 2 hadronic diboson searches from the ATLAS Collaboration~\cite{Aad:2015owa,ATLASJJ2015}. The full details of our procedure are given in appendix~\ref{sec:recast}. We generate dijet events at 8 and 13 TeV with {\scshape MadGraph5\_aMC@NLO}~\cite{Alwall:2014hca}, dividing the phase space in slices of 100 GeV of dijet invariant mass, from 700 GeV to 4 TeV. Event generation is followed by hadronisation and parton showering with {\scshape Pythia~8}~\cite{Sjostrand:2007gs}. Detector response is simulated with {\scshape Delphes 3}~\cite{deFavereau:2013fsa}, and {\scshape FastJet}~\cite{Cacciari:2011ma} is used to perform jet physics. For each slice of dijet invariant mass, $5\times 10^5$ events are simulated, and the results of each slice are weighed by the corresponding cross section and recombined to get the final (unnormalised) invariant mass distributions. Finally, a common scale factor is applied to all distributions within each run, so that the distributions without boson tagging are, to a good approximation, normalised to the measured ones for 20.3 fb$^{-1}$ in Run 1 and 3.2 fb$^{-1}$ in Run 2.

\begin{figure}[t]
\begin{center}
\begin{tabular}{ccc}
\includegraphics[height=5.2cm,clip=]{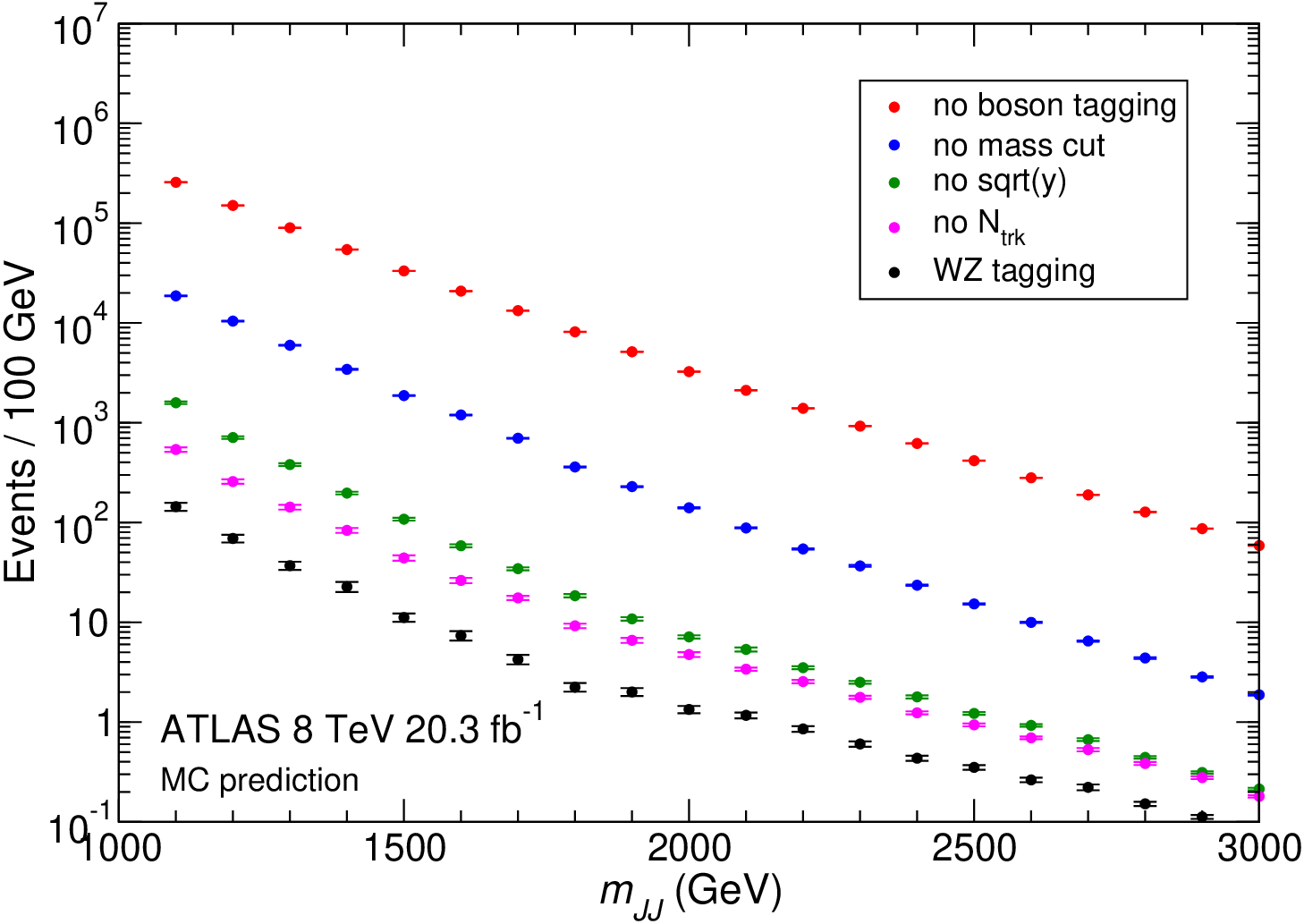} & \quad &
\includegraphics[height=5.2cm,clip=]{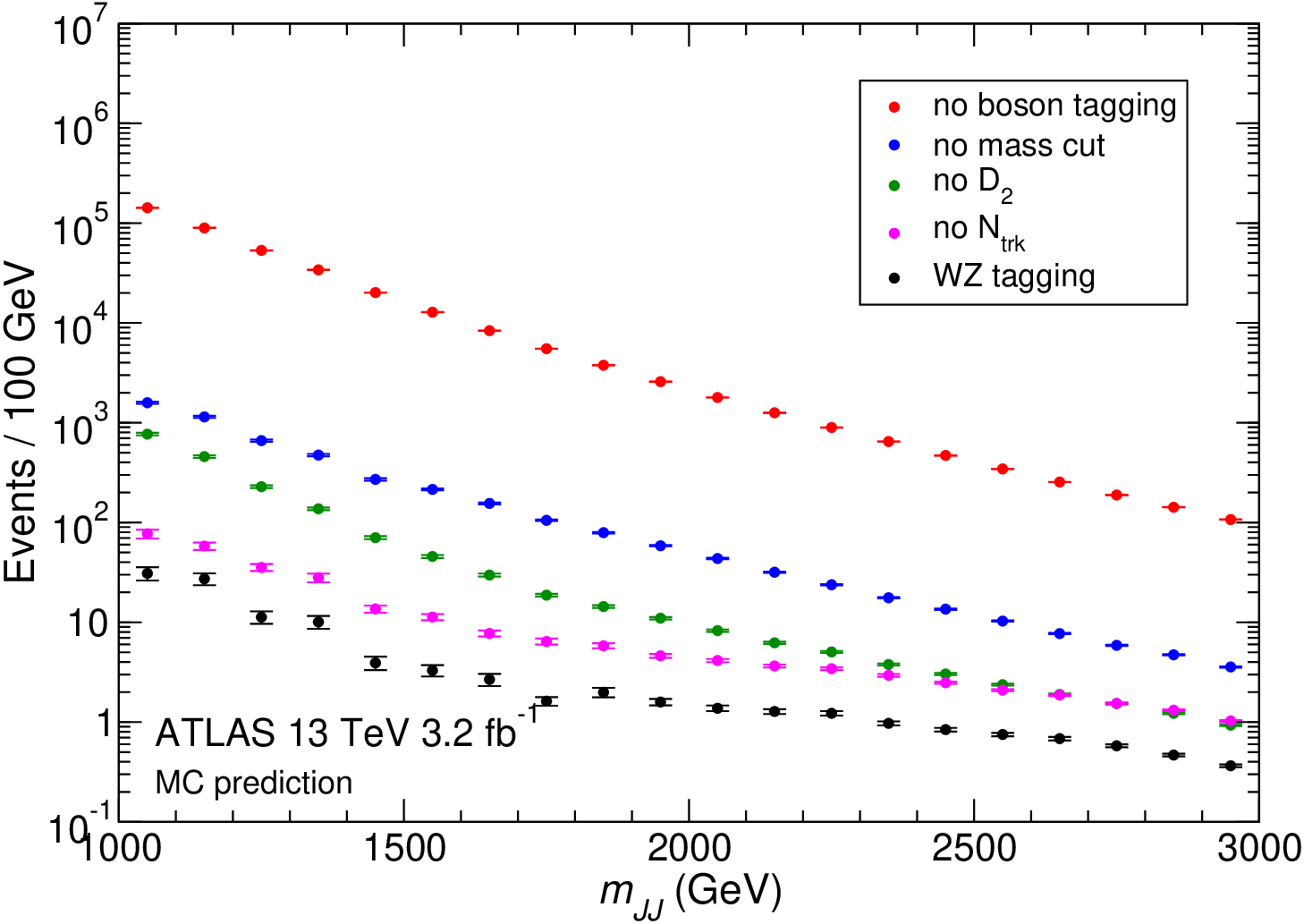}
\end{tabular}
\caption{Monte Carlo predictions for the QCD dijet background, in Run 1 (left) and Run 2 (right). The error bars in the points represent the Monte Carlo uncertainty.}
\label{fig:QCD}
\end{center}
\end{figure}

Our results are presented in figure~\ref{fig:QCD}. For Run 1, a slope decrease is visible around $1.7$ TeV in the distributions with full $WZ$ tagging (black) and without $\ntrk$ (pink). The appearance of these ``knees'' may cause that the apparent size of an excess near 2 TeV is two or three events larger than the actual size of the excess, and the extracted significance is consequently overestimated. For Run 2 there is also a slope decrease but milder and near $1.5-1.6$ TeV. The effect of the slope change on the significance of a bump at 2 TeV is expected to be much milder and likely absorbed in the fit. 

\begin{figure}[t]
\begin{center}
\includegraphics[height=5.2cm,clip=]{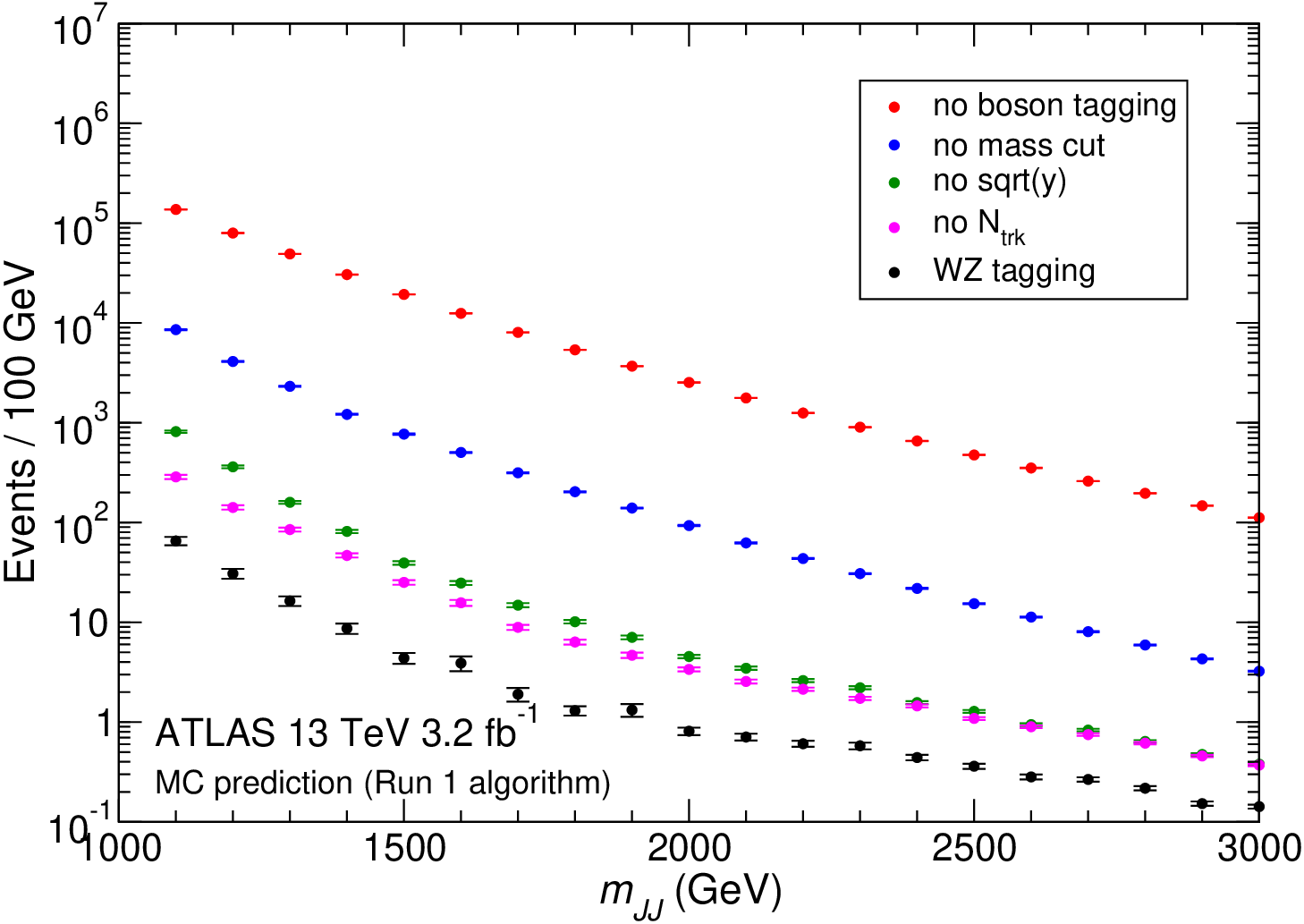}
\caption{Monte Carlo predictions for the QCD dijet background in Run 2, using the Run 1 analysis. The error bars in the points represent the Monte Carlo uncertainty.}
\label{fig:QCD2}
\end{center}
\end{figure}

It is also interesting to look at the background prediction for Run 2 but using the jet reconstruction and tagging of the Run 1 analysis. This is presented in figure~\ref{fig:QCD2}. We can observe that the change of slope near 1.7 TeV is similar to the one observed for the Run 1 analysis, namely figure~\ref{fig:QCD} (left). Therefore, apparently the differences between the ``knees'' observed in our dijet simulations for Run 1 and Run 2 are caused by the jet reconstruction and tagging, rather than the CM energy.

\newpage
\section{Triboson signals in diboson searches}
\label{sec:4}

In this section we explore benchmark triboson resonance scenarios and their potential signals in several final states. With this purpose, we fully recast several ATLAS and CMS diboson searches. The signals are generated at the parton level with {\scshape Protos}~\cite{protos}, using benchmark scenarios with $M_{W'} = 2.25$ TeV, $\Gamma_R = 50$ GeV, $M_Y = 650$ GeV, $\Gamma_Y = 5$ GeV, and (i) $X = H$; (ii) $X = A$, a light pseudo-scalar particle with mass $M_A \simeq M_Z$ and decays $A \to b \bar b$; (iii) $X = A$, with decays $A \to q \bar q$. For $Y^0$ the choice of the mass is obvious from the CMS 650 GeV excess~\cite{CMS:2016tio} and for $Y^\pm$ we take the same value for simplicity, despite the fact that their masses do not need to be the same~\cite{Aguilar-Saavedra:2015iew}. A heavier $X$ is also allowed, but if its mass is much larger than $M_Z$ the decays $Y^0 \to ZX$ cannot explain the CMS excess.
For each benchmark and CM energy a sample of $5 \times 10^4$ events is generated and passed through showering and detector simulation using different {\scshape Delphes} cards with settings adequate to the experimental analysis considered.
Additional details of our simulations are given in appendix~\ref{sec:recast}.

\subsection{ATLAS searches in the fully hadronic channel}

\begin{figure}[htb]
\begin{center}
\begin{tabular}{ccc}
\includegraphics[height=5.2cm,clip=]{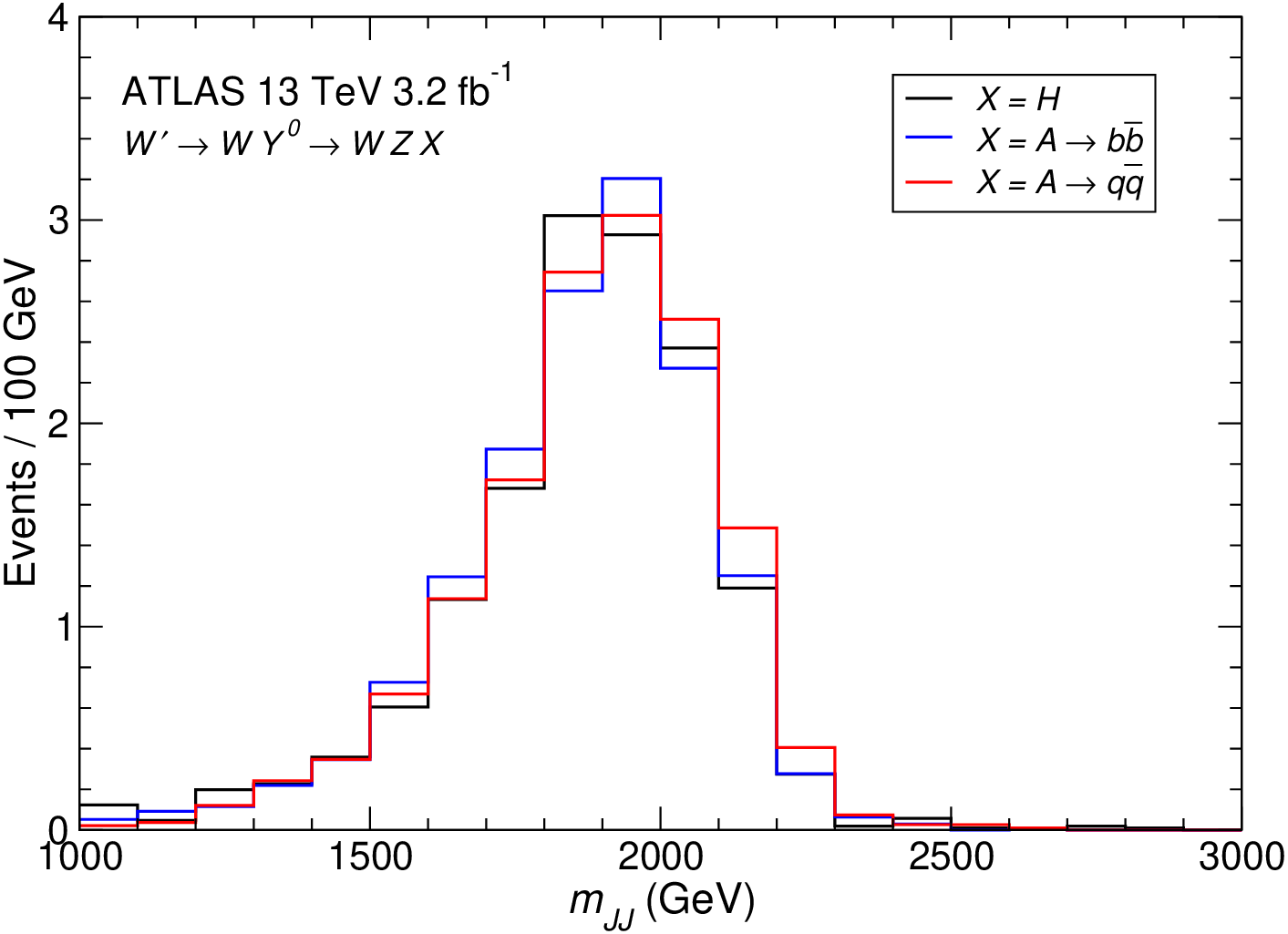} & \quad &
\includegraphics[height=5.2cm,clip=]{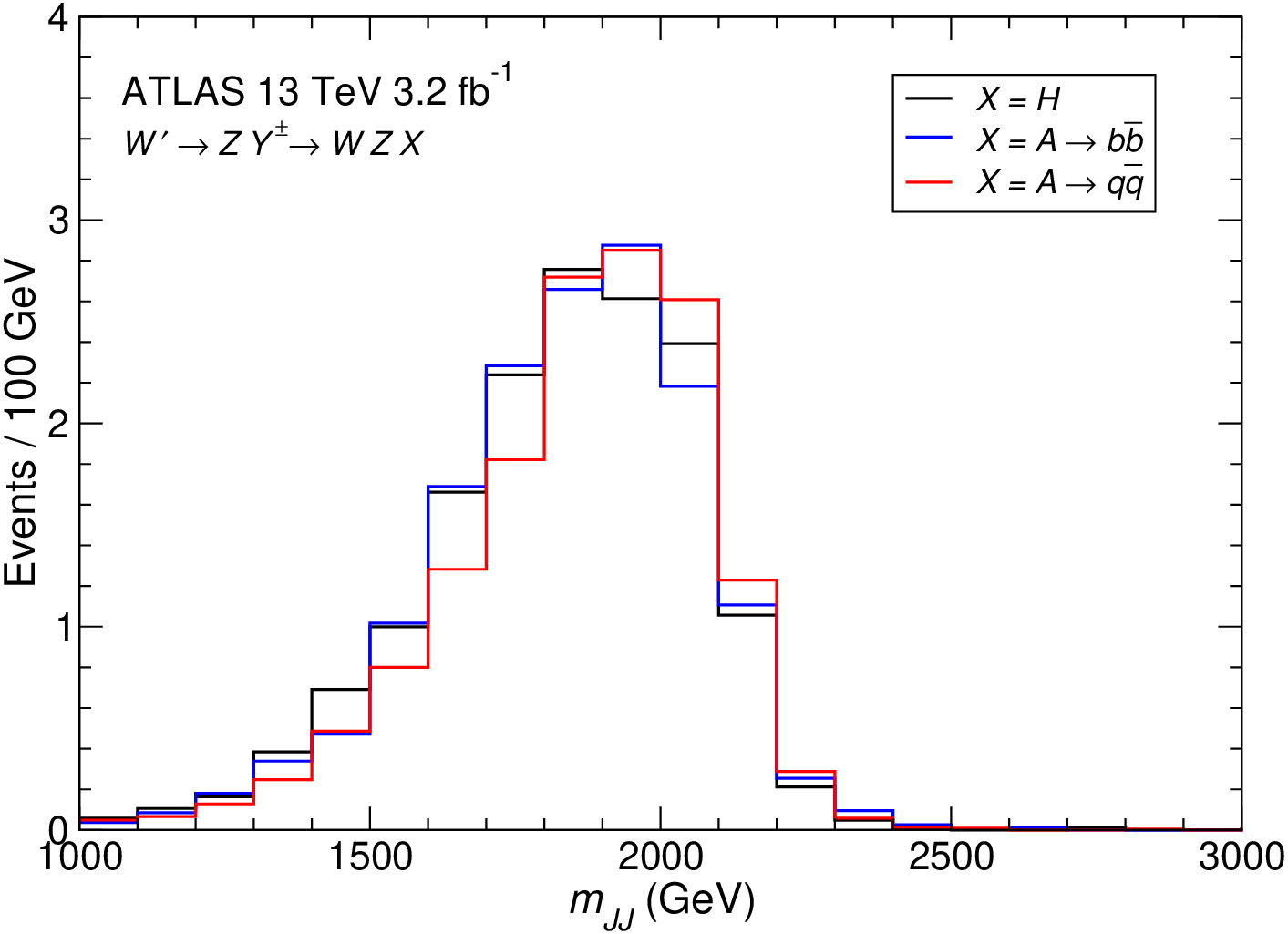} \\
\includegraphics[height=5.2cm,clip=]{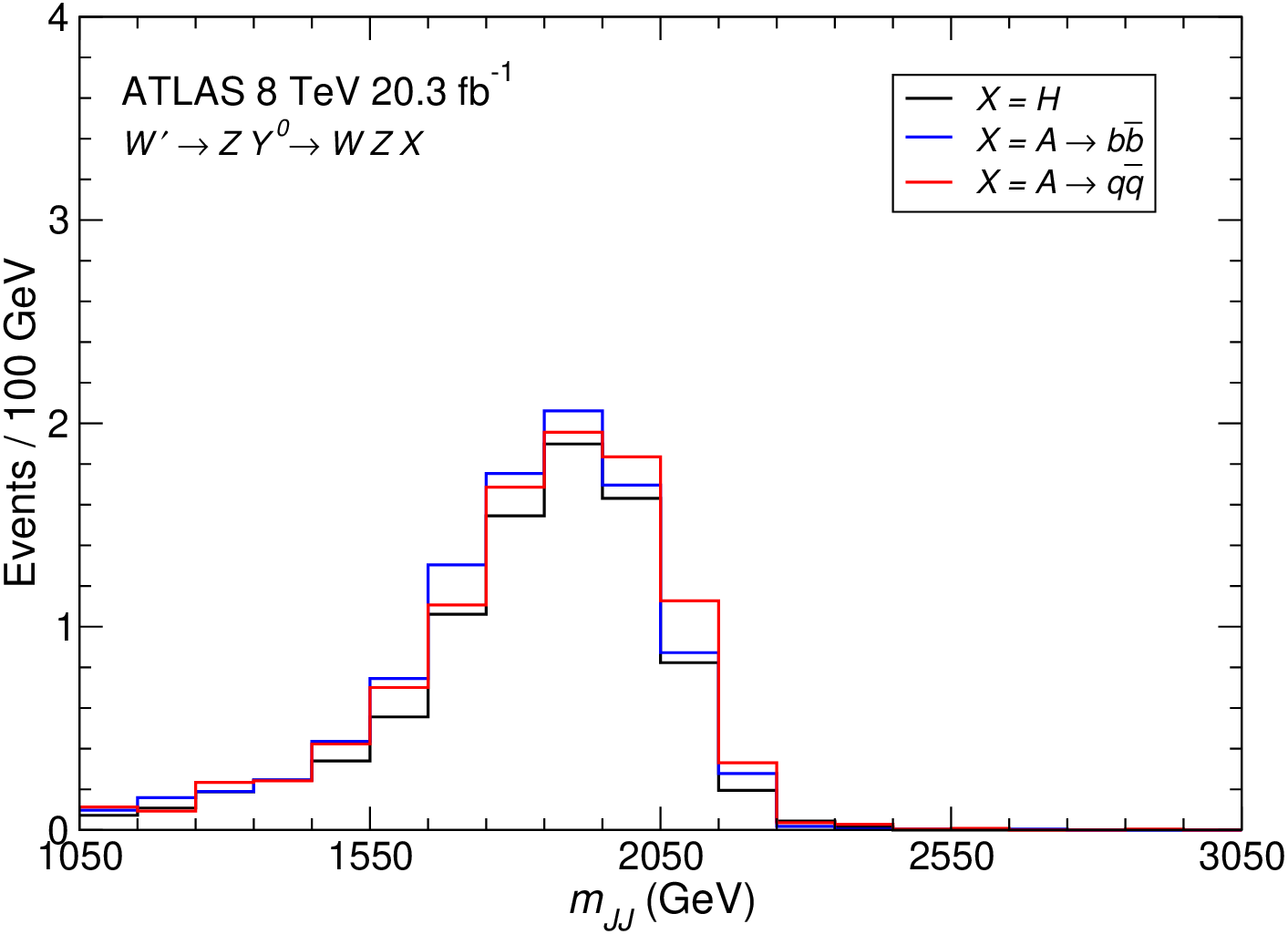} & \quad &
\includegraphics[height=5.2cm,clip=]{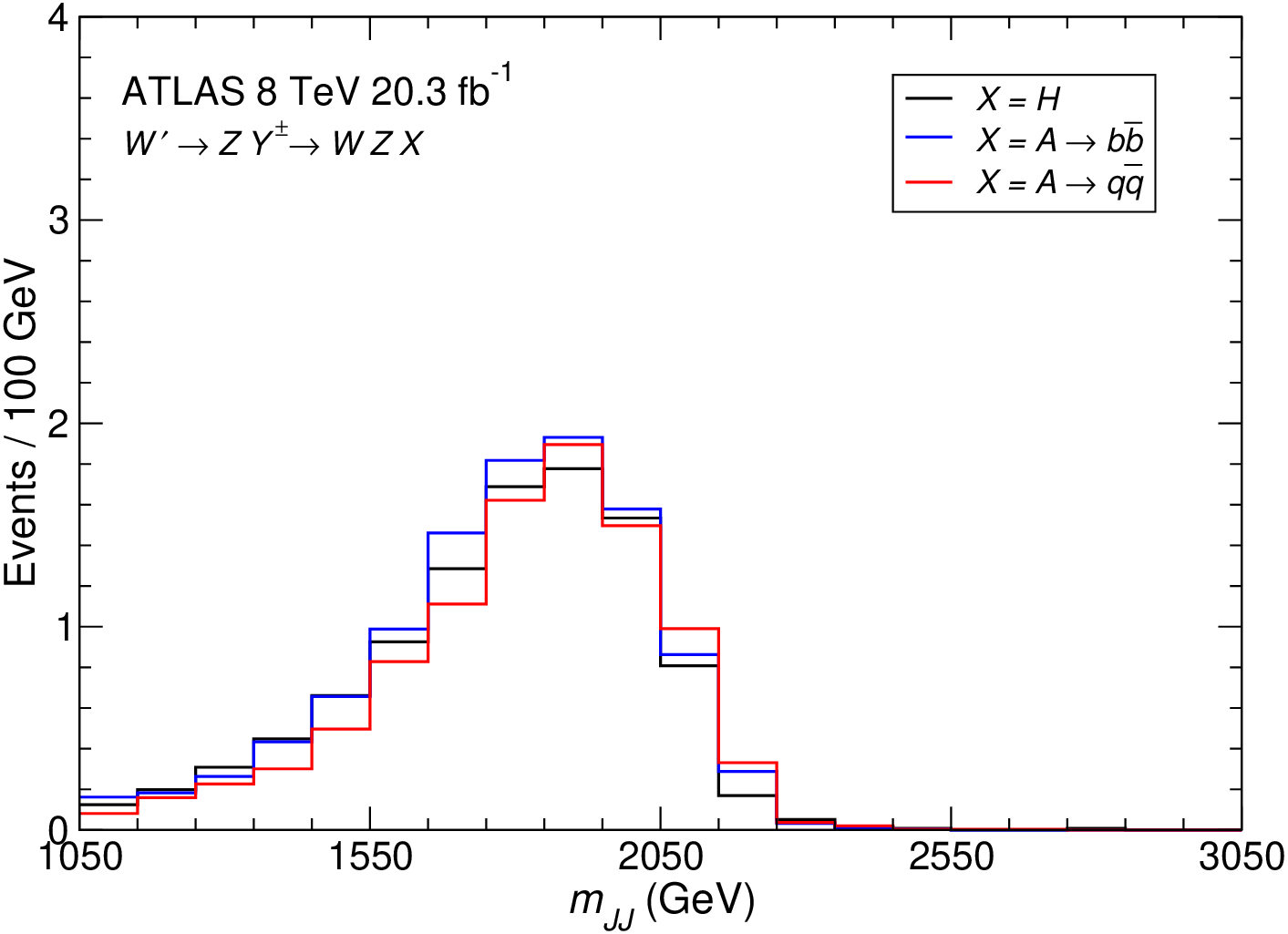}
\end{tabular}
\caption{Top: Dijet invariant mass distribution for triboson signals in the ATLAS Run 2 fully hadronic analysis~\cite{ATLASJJ2015}, in the $WZ$ selection without $\ntrk$. Bottom: the same, for the Run 1 analysis~\cite{Aad:2015owa}.}
\label{fig:A2}
\end{center}
\end{figure}

We begin with the Run 2 analysis focusing on the $WZ$ selection without the $\ntrk$ cut, and use the excess of events over the background-only expectation in the analysis of ref.~\cite{ATLASJJ2015} to fix the overall normalisation of our potential signals in all channels. Specifically, we require 10 signal events in the invariant mass interval $1.7 - 2.1$ TeV. The $m_{JJ}$ distributions with this normalisation for the two possible $W'$ cascade decay channels, $W' \to W Y^0 \to WZX$, $W' \to Z Y^\pm \to W Z X$, are presented in figure~\ref{fig:A2} (top). In all cases, the distributions have a peaked shape, in agreement with earlier results~\cite{Aguilar-Saavedra:2015rna} obtained with a much less elaborate simulation. We point out that the peaked shape of the dijet invariant mass distribution for triboson signals in the ATLAS analyses is rather independent of the mass of the intermediate particle, and results are quite similar in this respect for masses from 300 GeV to 1 TeV. The efficiencies found for these channels are collected in table~\ref{tab:A2}, for neutral ($Y^0$) and charged ($Y^\pm$) intermediate particles. These efficiencies are comparable to the efficiency of 0.09 found for a $WZ$ diboson signal without $\ntrk$. The efficiency penalty of $1/3-2/3$ for triboson signals is much smaller than the value of $1/7$ estimated in ref.~\cite{Aguilar-Saavedra:2015rna} with a more simplistic analysis, because kinematical configurations where two of the bosons merge into a single jet, which were discarded there, may also pass the event selection criteria due to the filtering performed on the jets. As a consequence, the coupling of the $W'$ boson eventually required to explain the size of the excess is not too large and remains perturbative (see section~\ref{sec:6}).

\begin{table}[htb]
\begin{center}
\begin{tabular}{cccc}
& $X = H$ & $X = A \to b \bar b$ & $X = A \to q \bar q$ \\
$Y^0$ & 0.030 & 0.050 & 0.055 \\
$Y^\pm$ & 0.032 & 0.058 & 0.066 \\
\end{tabular}
\caption{Efficiencies for triboson signals, relative to the full $WZX$ samples with all possible decays, for the ATLAS Run 2 fully hadronic analysis~\cite{ATLASJJ2015}, in the $WZ$ selection without $\ntrk$.}
\label{tab:A2}
\end{center}
\end{table}

For the Run 1 ATLAS analysis with 20.3 fb$^{-1}$ we obtain the distributions in figure~\ref{fig:A2} (bottom) for the $WZ$ selection without the $\ntrk$ cut. The same signal normalisations are used, with a scaling factor of $1/8.3$ for the cross section at 8 TeV with respect to 13 TeV (for $M_{W'} = 2.25$ TeV), and the corresponding luminosity scaling.
The predicted size of the signals is smaller than in the Run 2 analysis, $3-4$ events fewer at the $1.7-2.1$ TeV invariant mass interval, due to two effects: first, the efficiencies, collected in table~\ref{tab:A1}, are slightly smaller; second, for a $W'$ mass of 2.25 TeV the larger luminosity does not compensate the smaller cross section at 8 TeV. As we have seen in section~\ref{sec:2}, the fitted excess in Run 1 is larger than in Run 2; however, as argued in section~\ref{sec:3}, a knee in the background distribution could cause an apparent excess of a few events as well. Furthermore, in a sample of around 10 events one also expects statistical fluctuations to have some relevance.

\begin{table}[t]
\begin{center}
\begin{tabular}{cccc}
& $X = H$ & $X = A \to b \bar b$ & $X = A \to q \bar q$ \\
$Y^0$ & 0.025 & 0.046 & 0.051 \\
$Y^\pm$ & 0.028 & 0.054 & 0.059 \\
\end{tabular}
\caption{Efficiencies for triboson signals, relative to the full $WZX$ samples with all possible decays, for the ATLAS Run 1 fully hadronic analysis \cite{Aad:2015owa}, in the $WZ$ selection without $\ntrk$.}
\label{tab:A1}
\end{center}
\end{table}

\subsection{CMS searches in the fully hadronic channel}

\begin{figure}[htb]
\begin{center}
\begin{tabular}{ccc}
\includegraphics[height=5.2cm,clip=]{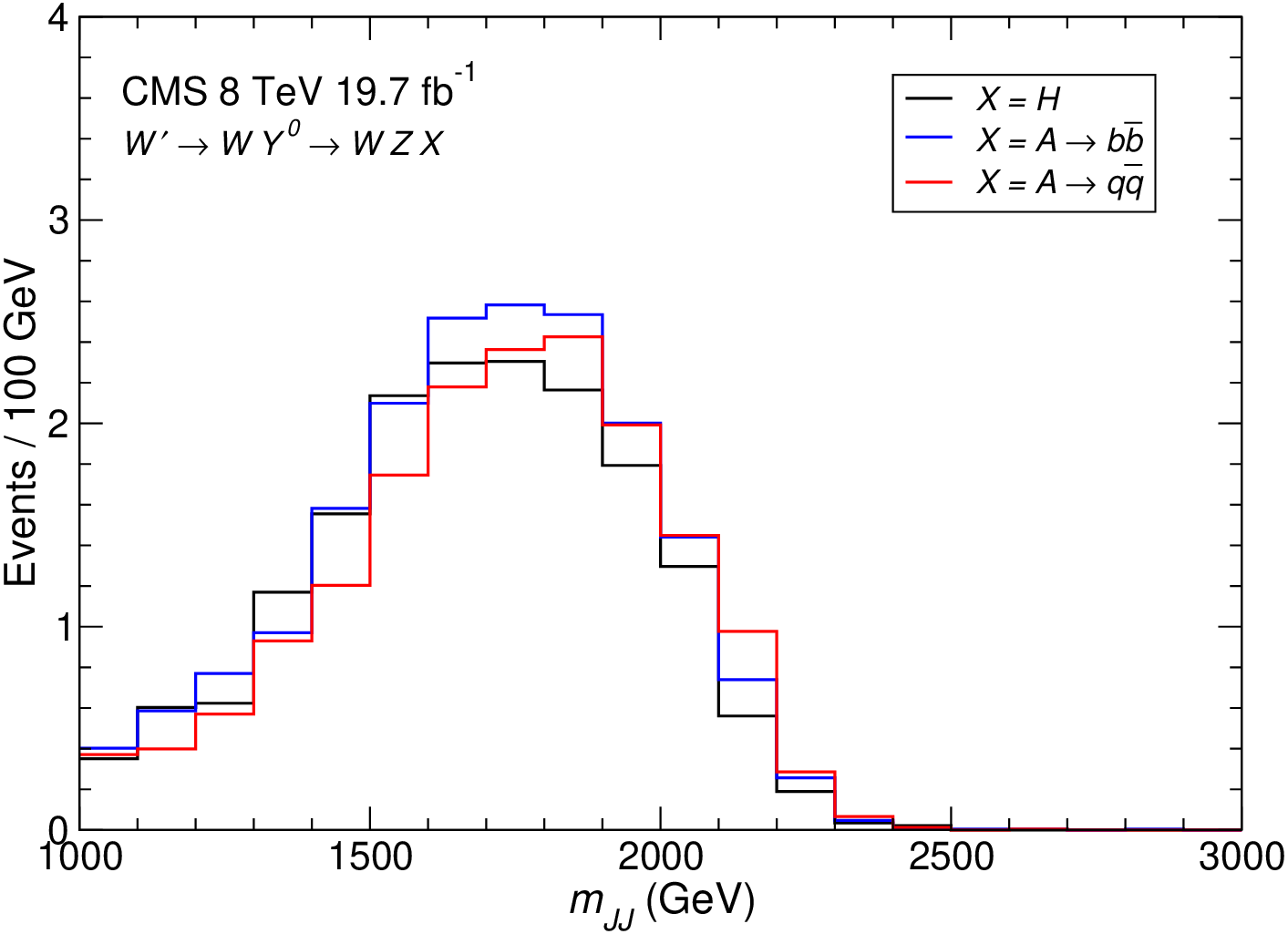} & \quad &
\includegraphics[height=5.2cm,clip=]{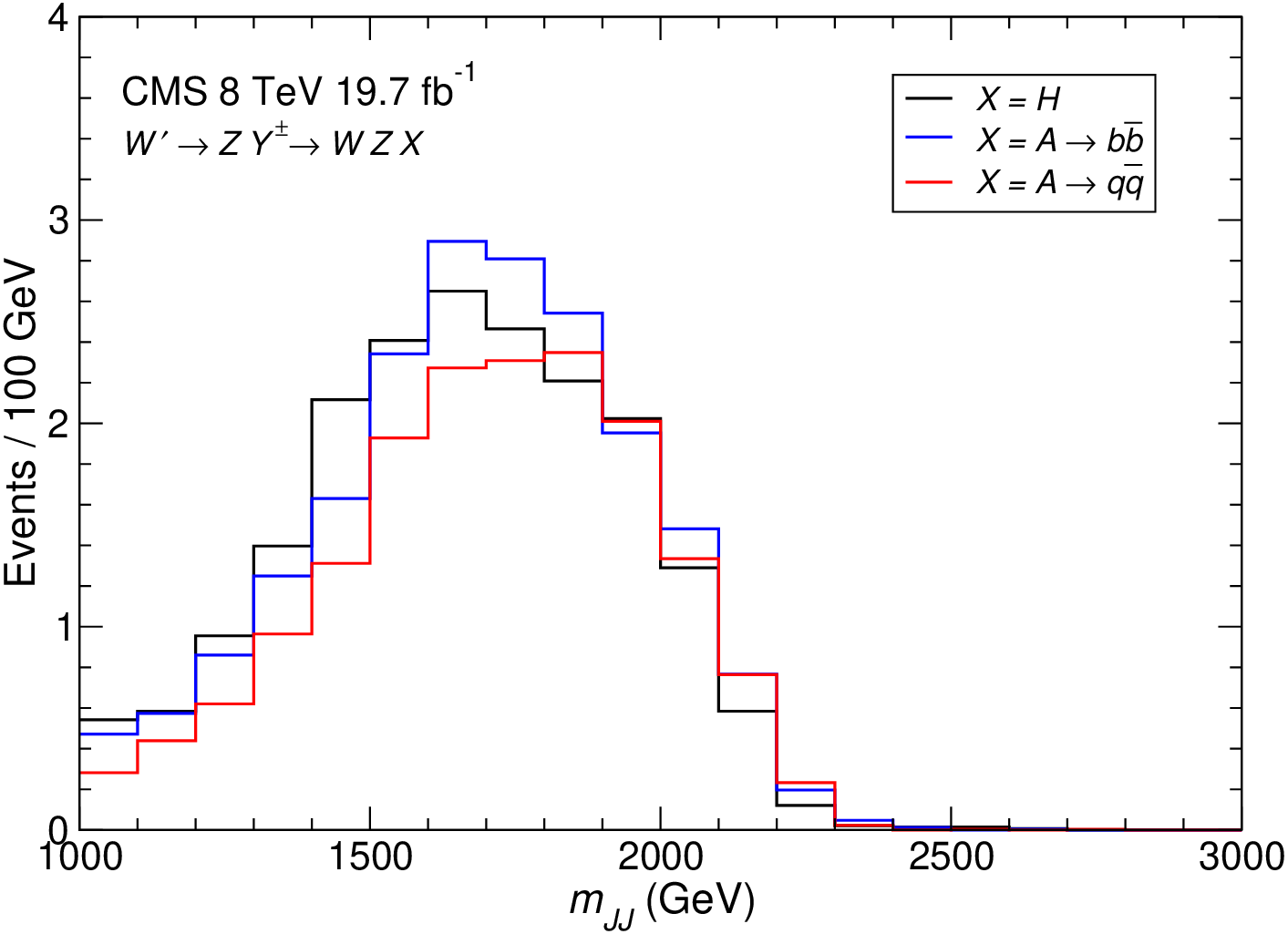} \\
\includegraphics[height=5.2cm,clip=]{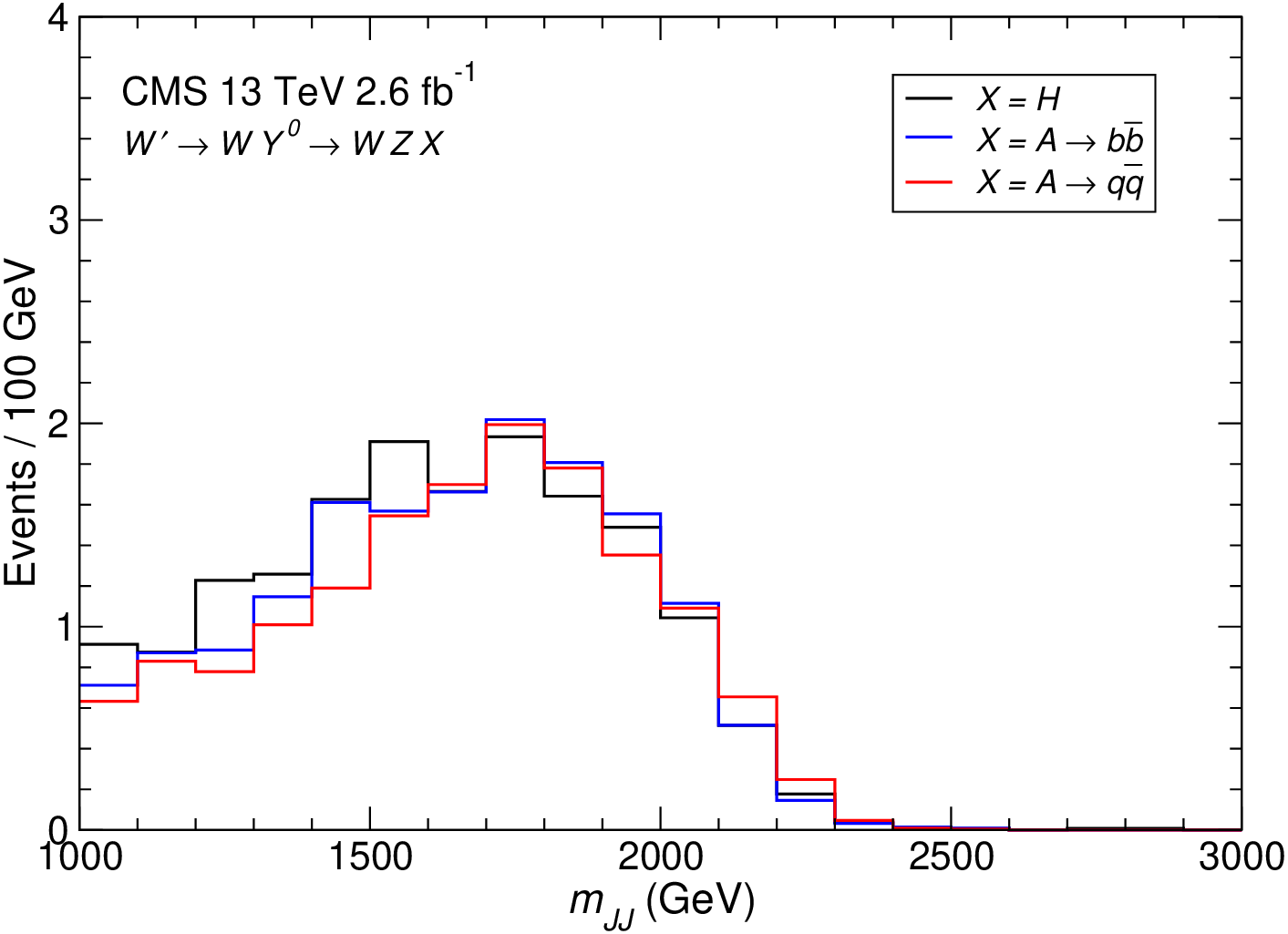} & \quad &
\includegraphics[height=5.2cm,clip=]{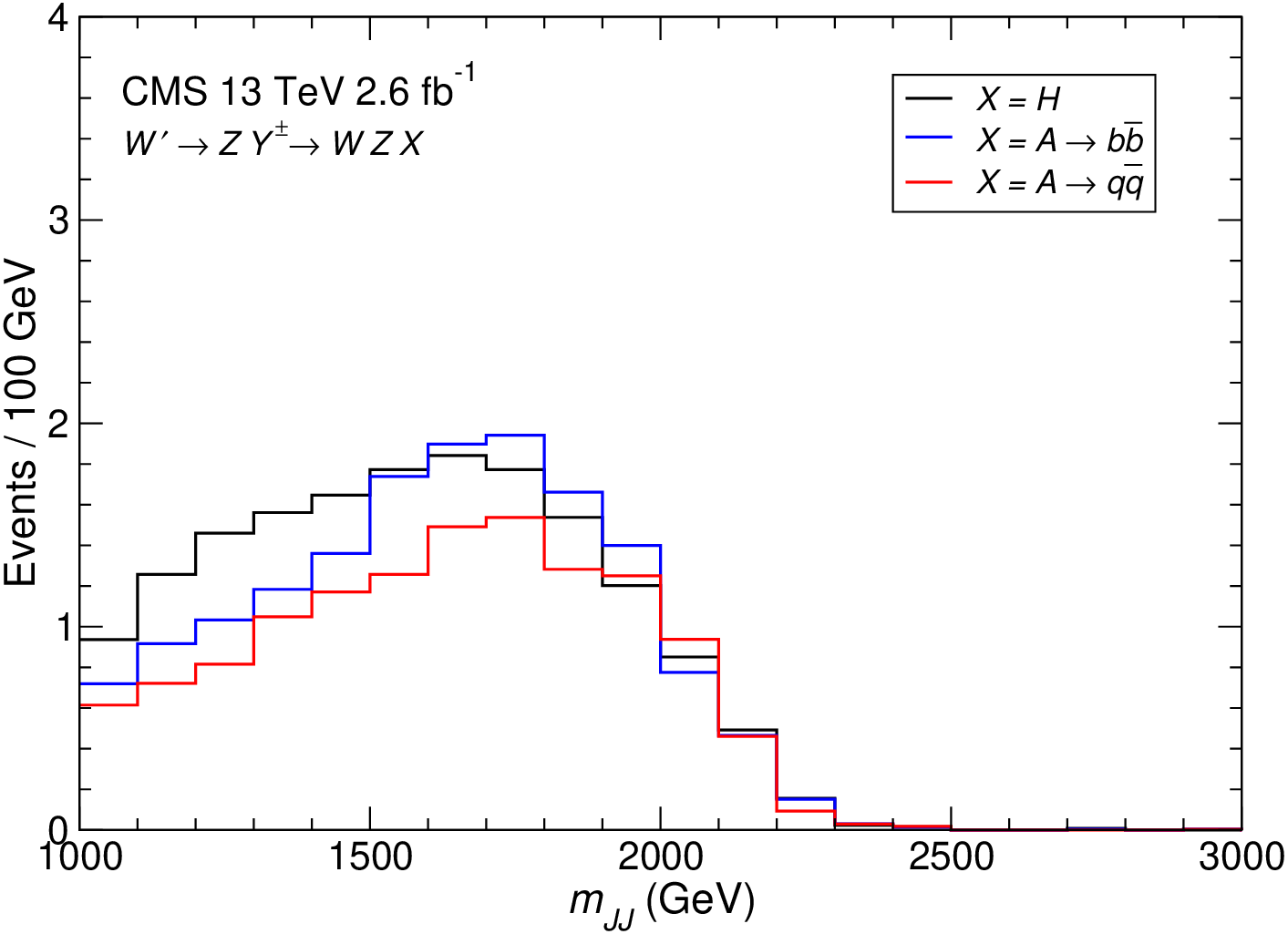}
\end{tabular}
\caption{Dijet invariant mass distribution for triboson signals in the CMS Run 1~\cite{Khachatryan:2014hpa} (top) and Run 2~\cite{CMS:2015nmz} (bottom) fully hadronic analyses, in the high-purity sample.}
\label{fig:CMS12}
\end{center}
\end{figure}

The search for diboson resonances decaying into two fat jets performed by the CMS Collaboration with 8 TeV data and a luminosity of 19.7 fb$^{-1}$~\cite{Khachatryan:2014hpa} showed a small excess, near the $2\sigma$ level, in the $1.5 - 2$ TeV range, whereas the search at 13 TeV with 2.6 fb$^{-1}$, with a similar sensitivity for 2 TeV resonances, is compatible with the SM expectation at the $1\sigma$ level. We give our results for the CMS Run 1 analysis in figure~\ref{fig:CMS12} (top), considering the high-purity sample. (The signal efficiency for the low-purity sample is similar and the background much larger.) In this analysis, the discrimination between $W$ and $Z$ bosons is not attempted, and jet masses in the interval $70-100$ GeV are considered. We use the same signal normalisations as before, with a scale factor of $1/8.3$ for 8 TeV cross sections. In contrast to the ATLAS searches, here the dijet invariant mass distributions of the triboson signals are wider, in the $1.5 - 2$ TeV range, with around two events per 100 GeV. These predictions are compatible with the measurements, especially bearing in mind that the QCD background itself is normalised from measured data.

In the Run 2 search the CMS Collaboration follows a slightly different strategy and divides the boson-tagged dijet samples into $WW$, $WZ$ and $ZZ$, where a jet is considered as $W$-tagged if its mass 
is in the range $65-85$ GeV, and $Z$-tagged for a mass in the range $85-105$ GeV. We give in figure~\ref{fig:CMS12} (bottom) the predictions for the high-purity $WZ$-tagged sample, which amount to $1-2$ signal events per 100 GeV. These predictions are also compatible with the measurements, where a handful of extra events above the SM prediction, as well as some downward fluctuations, are seen over the $1.5-2$ TeV range in the different $WW$, $WZ$ and $ZZ$ samples.

\subsection{ATLAS Run 2 search in the $\nu\nu J$ channel}
\label{sec:vvJ}

\begin{figure}[htb]
\begin{center}
\begin{tabular}{ccc}
\includegraphics[height=5.2cm,clip=]{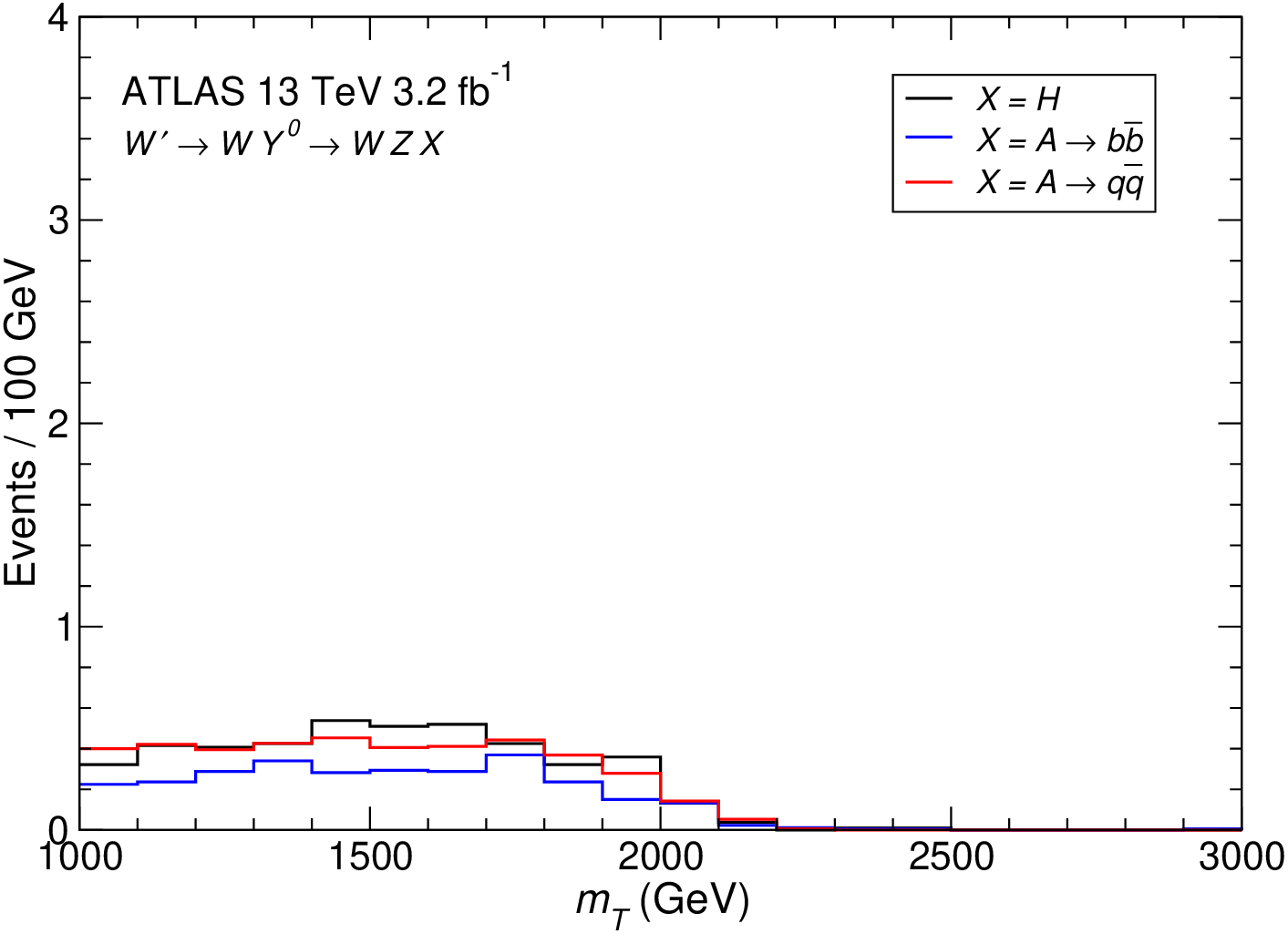} & \quad &
\includegraphics[height=5.2cm,clip=]{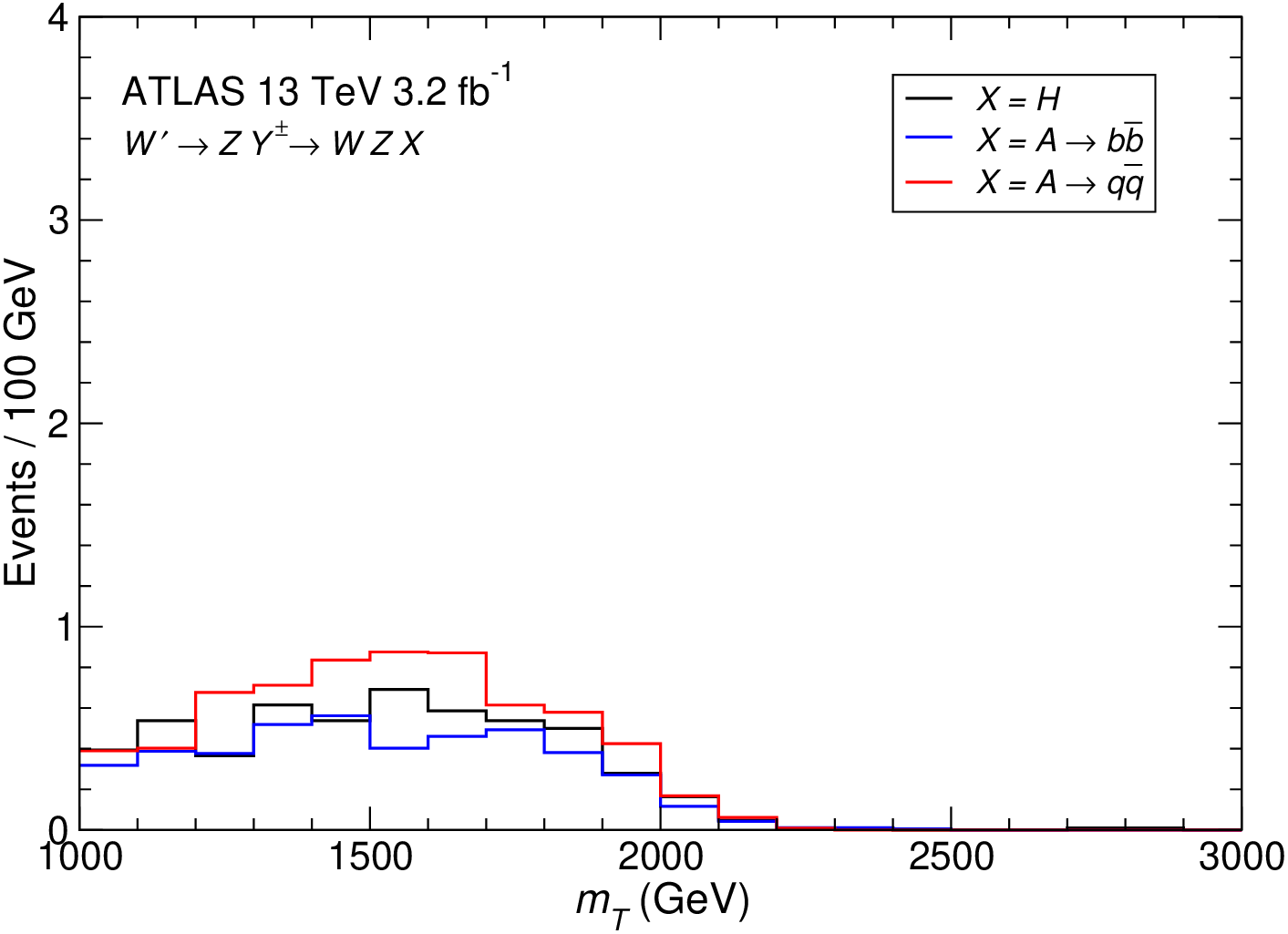}
\end{tabular}
\caption{Transverse mass distribution for triboson signals in the ATLAS Run 2 $\nu\nu J$ analysis~\cite{ATLASvvJ2015}, in the $WZ$ selection.}
\label{fig:vvJ}
\end{center}
\end{figure}

The $\nu \nu J$ channel (i.e. a final state with a single fat jet plus large missing energy) provides the best sensitivity to diboson resonances among the searches performed by the ATLAS Collaboration using Run 2 data~\cite{ATLASvvJ2015}. In this search the diboson mass cannot be directly reconstructed and, instead, the event transverse mass $m_T$ is considered. It turns out that the sensitivity of this channel to triboson resonances is very poor, since the event selection requires the missing transverse energy vector to be isolated from any other jets (which is infrequent in the case of a triboson resonance) and also vetoes any charged lepton, often present in the final state under consideration. For these reasons, the sensitivity to triboson resonances is a factor $10-15$ worse than in the fully hadronic channel. We present our results for this analysis 
in figure~\ref{fig:vvJ}, with the same signal normalisation used in previous examples. The expectation for a signal that reproduces the ATLAS Run 2 dijet excess is one or at most two events in the $1.5-2$ TeV range, perfectly compatible with the null results of this search.

\subsection{ATLAS Run 2 search in the $\ell\nu J$ channel}

\begin{figure}[htb]
\begin{center}
\begin{tabular}{ccc}
\includegraphics[height=5.2cm,clip=]{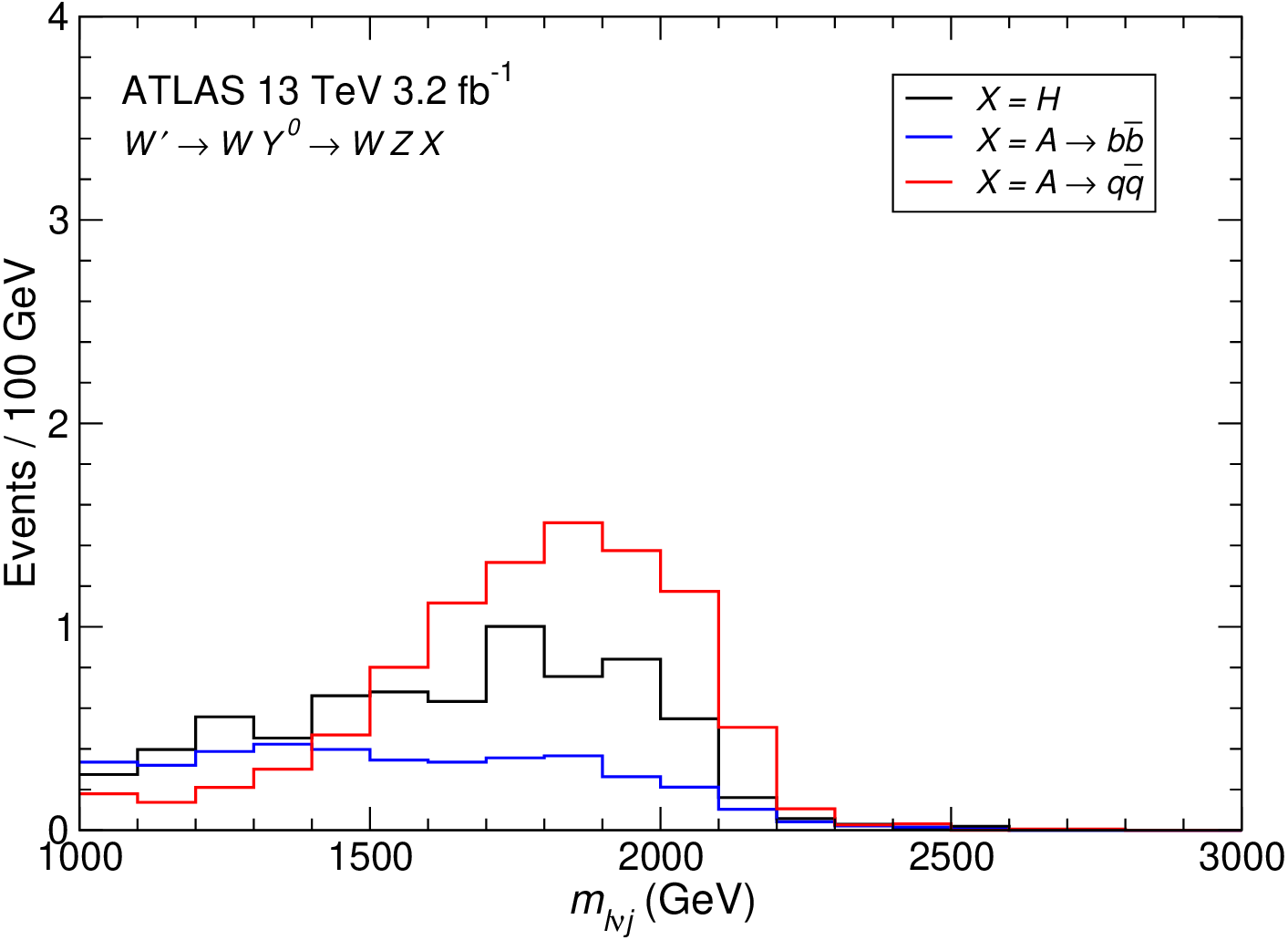} & \quad &
\includegraphics[height=5.2cm,clip=]{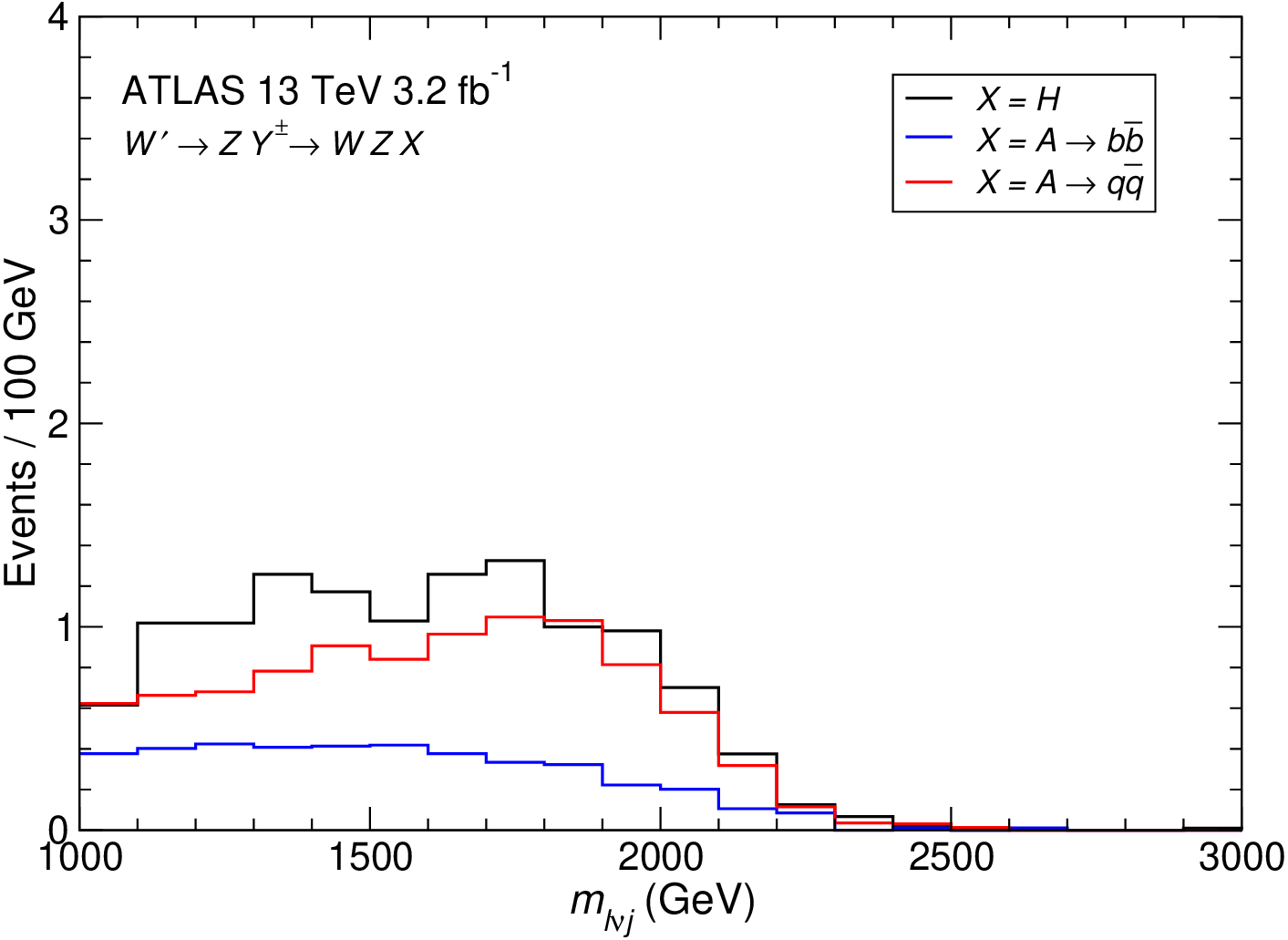}
\end{tabular}
\caption{Reconstructed $\ell \nu J$ mass distribution for triboson signals in the ATLAS Run 2 $\ell \nu J$ analysis~\cite{ATLASlvJ2015}, in the $WZ$ selection.}
\label{fig:lvJ}
\end{center}
\end{figure}

This channel~\cite{ATLASlvJ2015,CMS:2015nmz}, in which the final state has a charged lepton $\ell$, large missing energy and a fat jet, is also very sensitive to diboson resonances --- slightly more than the fully hadronic channel --- but much less sensitive to triboson resonances. First, because the would-be reconstructed diboson mass, $m_{\ell \nu J}$, does not peak at the $WZX$ invariant mass for a triboson resonance, but is much broader instead. And, especially, because the event selection implemented by both the ATLAS and CMS Collaborations include a veto on $b$-tagged jets in order to reduce the background from $t \bar t$ production, and this veto suppresses the signals where $X$ has dominant decay into $b \bar b$. We restrict ourselves to the ATLAS analysis, as the CMS one follows a similar strategy, and present our results, conveniently normalised as in the previous cases, in figure~\ref{fig:lvJ}. We consider the $WZ$ selection, that is, a window of $\pm 13$ GeV around the expected $Z$ mass peak for the jet mass $m_J$.

The two benchmark models with $X \to q \bar q$ seem to be disfavoured by the null result of the ATLAS search~\cite{ATLASlvJ2015}, as is the benchmark with $W' \to Y^\pm Z  \to WZH$. However, the caveat is that the normalisation of the SM background in this analysis is determined from a fit using two control regions:
\begin{itemize}
\item[(i)] the $W$ control region, with the same event selection as for the signal region but inverting the jet mass requirement: $50 < m_J < 70.2$ or $m_J > 106.4$ GeV;
\item[(ii)] the top control region, with the same event selection as for the signal region but requiring a $b$-tagged jet instead of vetoing them.
\end{itemize}
For illustration, we collect in table~\ref{tab:lvJ} the efficiency for triboson signals in the $W$ and top control regions, referred to the efficiency for the $WZ+WW$ signal region. 
The number of data events in the $m_{\ell \nu J}$ distributions for the signal and top control regions is of the same order, and it is around 5 times larger for the $W$ control region. From this exercise we can conclude that the benchmarks with $X = A \to q \bar q$ indeed seem disfavoured by data while the rest, especially for $A \to b \bar b$, are compatible with the current limits.

\begin{table}[htb]
\begin{center}
\begin{tabular}{cccc}
               &     $X = H$      &    $X = A \to b \bar b$ & $X = A \to q \bar q$ \\
$Y^0$     & $1:1.20:0.55$ & $1:0.40:0.90$ & $1:0.40:0.10$ \\
$Y^\pm$ & $1:0.55:0.75$ & $1:0.35:1.60$ & $1:0.35:0.05$ \\
\end{tabular}
\caption{Efficiencies $\epsilon$ for triboson signals in the $W$ and top control regions, relative to the efficiency in the signal region, using the notation $1 \;:\; \epsilon_W \;:\; \epsilon_\text{top}$}
\label{tab:lvJ}
\end{center}
\end{table}

\subsection{CMS Run 2 search in the $\ell\ell J$ channel}
\label{sec:CMSllJ}

\begin{figure}[t]
\begin{center}
\begin{tabular}{ccc}
\includegraphics[height=5.2cm,clip=]{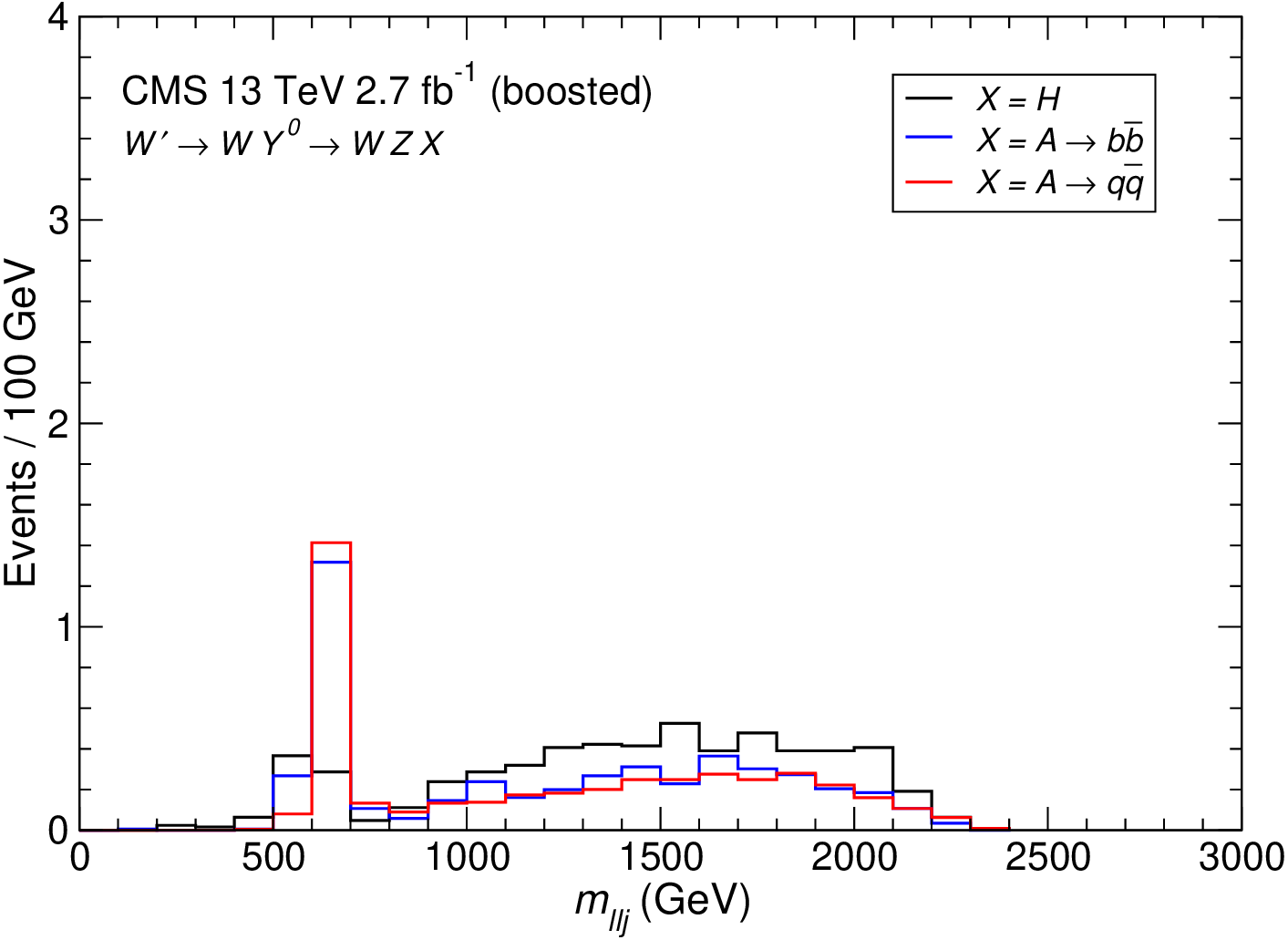} & \quad &
\includegraphics[height=5.2cm,clip=]{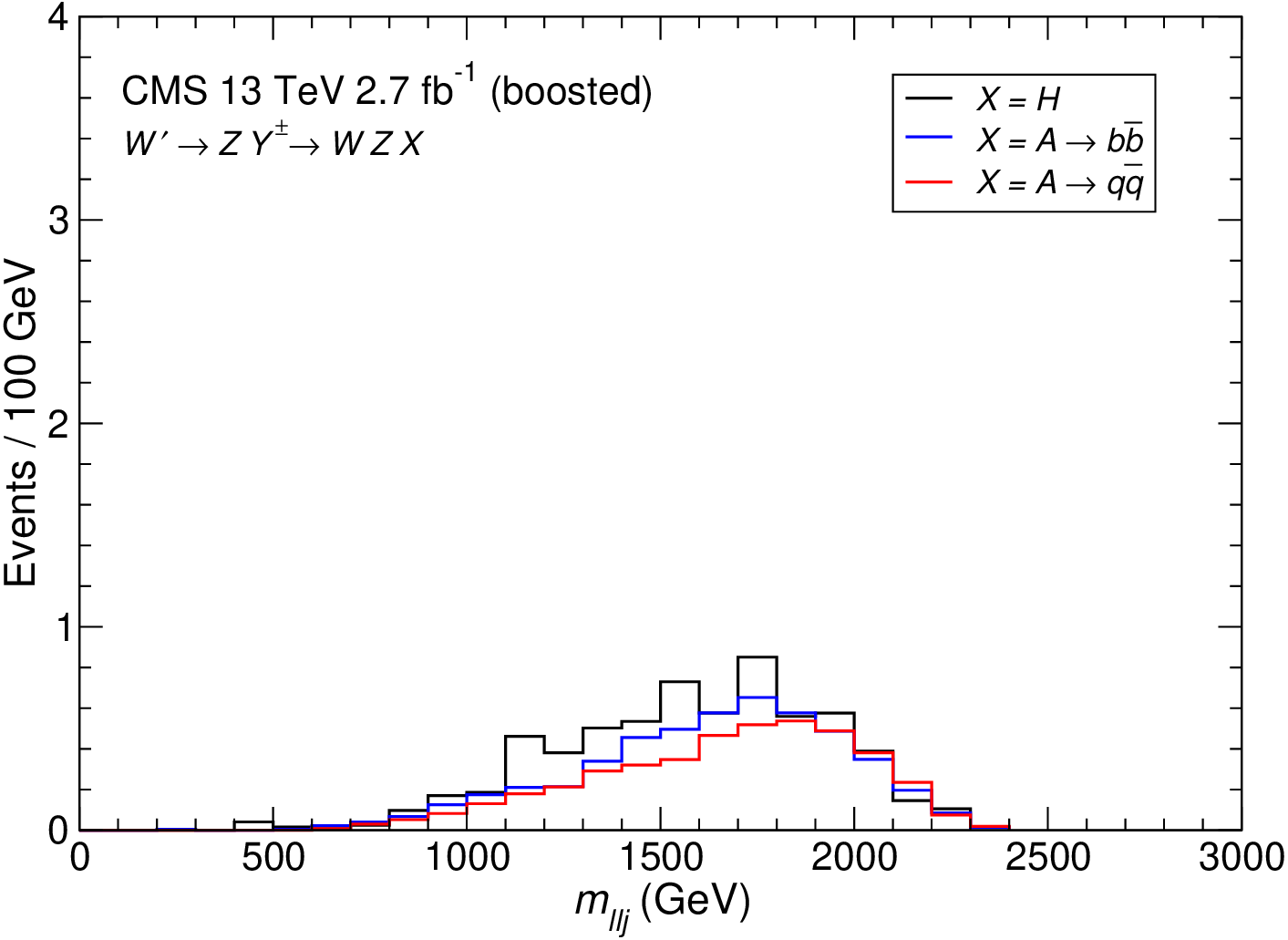} \\
\includegraphics[height=5.2cm,clip=]{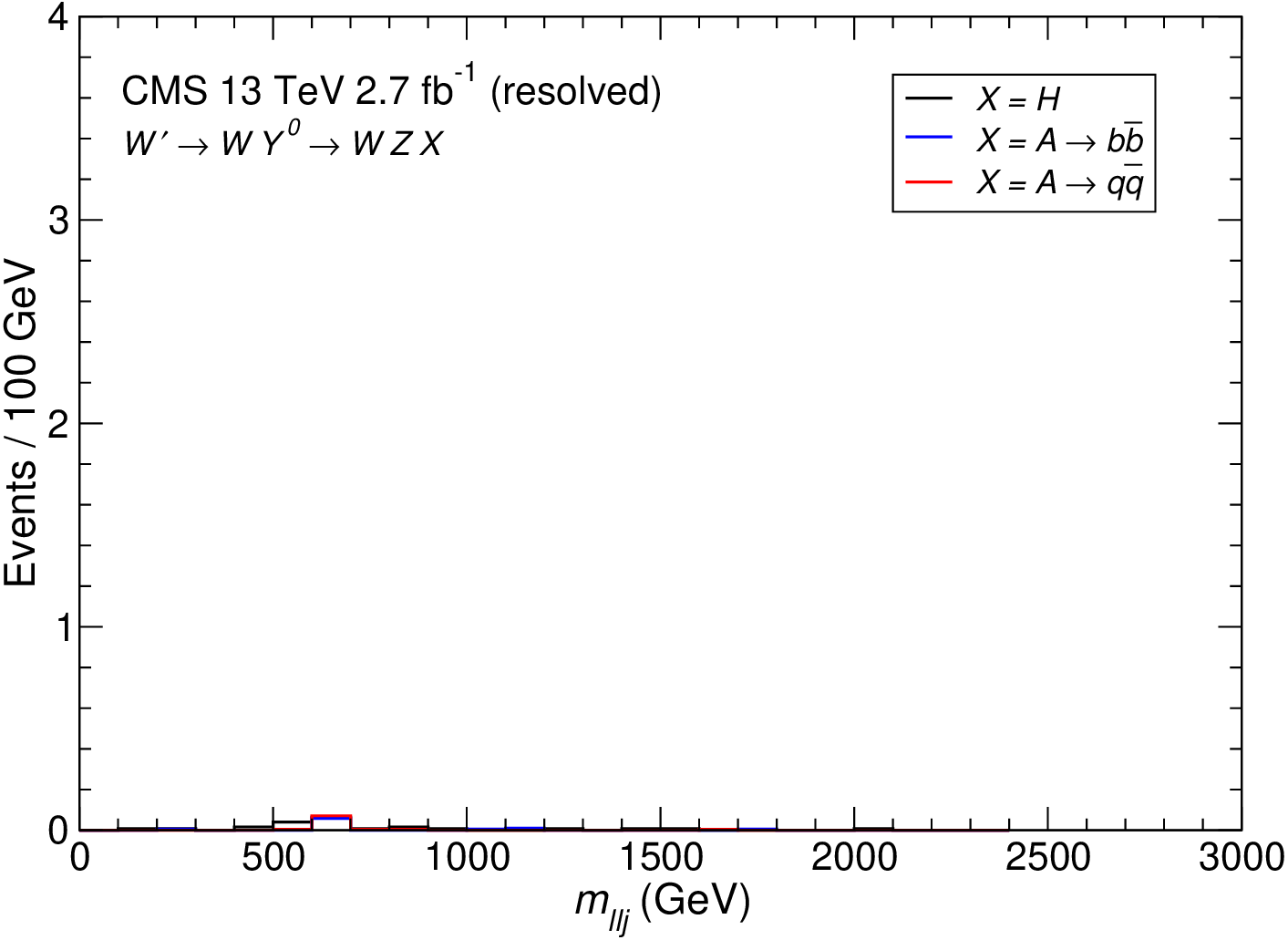} & \quad &
\includegraphics[height=5.2cm,clip=]{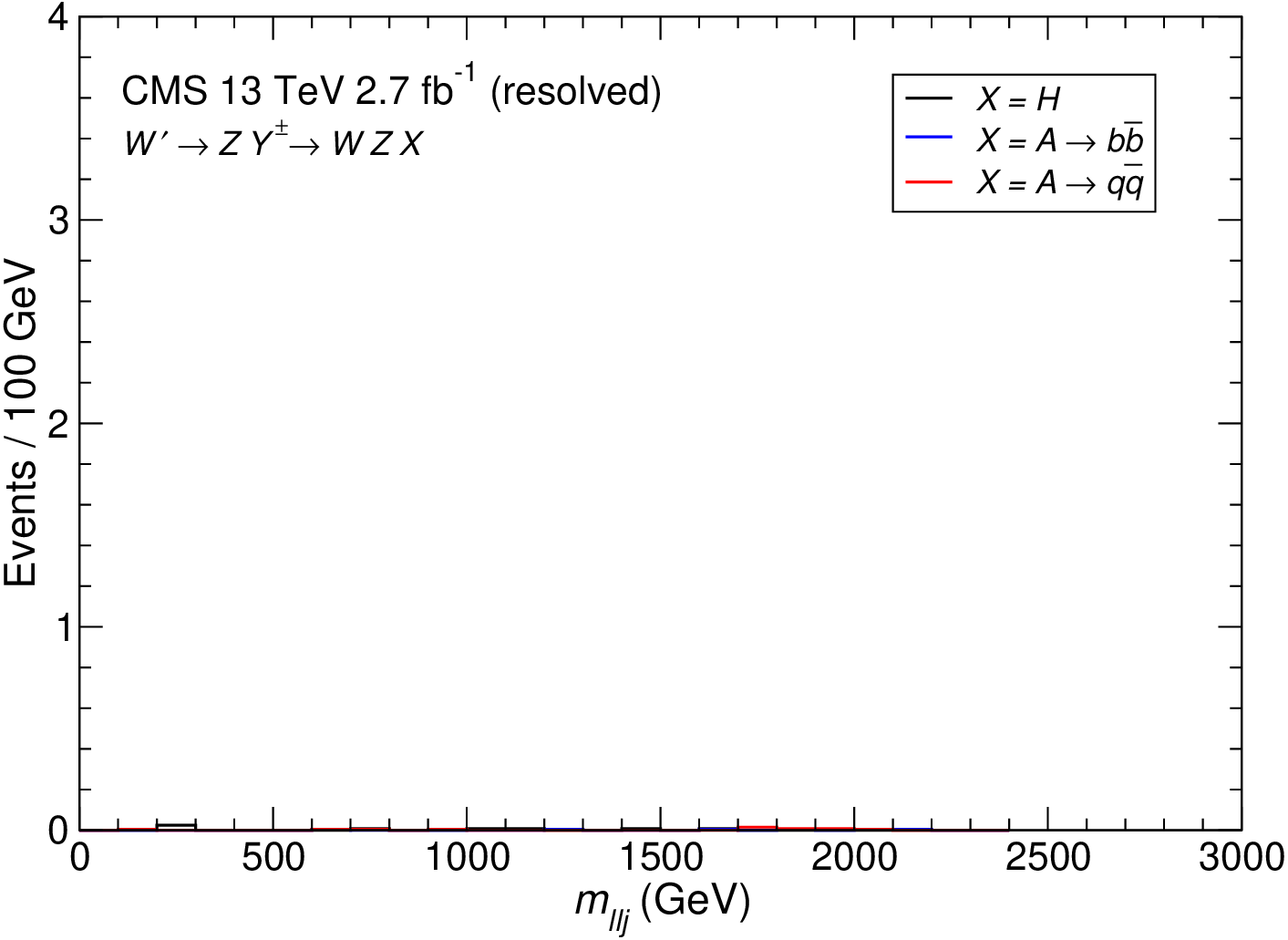}
\end{tabular}
\caption{Top: reconstructed $\ell \ell J$ mass distribution for triboson signals in the CMS Run 2 $\ell \ell J$ low-mass boosted analysis~\cite{CMS:2016tio}. Down: the same, for the low-mass resolved analysis.}
\label{fig:llJ_CMS}
\end{center}
\end{figure}

Diboson searches in this final state are performed by looking at a peak in the invariant mass distribution of two oppositely-charged leptons and a boson-tagged jet. Triboson signals do not exhibit a peak in the $\ell \ell J$ invariant mass distribution near the heavy resonance mass $M_{R} \simeq 2$ TeV, but a peak may appear --- if the event selection looking for high-mass resonances does not suppress it --- at the $Y^0$ mass, resulting from the decay $Y^0 \to Z X$. The CMS Collaboration performs two separate searches in this channel with Run 2 data~\cite{CMS:2016tio}: one aiming for low-mass resonances with masses $0.55- 1.4$ TeV, and a second one for resonances in the range $0.8 - 2.5$ TeV. We will focus here in the low-mass search, and for the high-mass range we will consider the ATLAS search~\cite{ATLASllJ2015}. The CMS low-mass search includes two channels, the boosted channel where a high-purity large-$R$ ($R=0.8$) fat jet with mass $65~\text{GeV} < m_J < 105~\text{GeV}$ is found, in which case it is the candidate for the vector boson $V$, and the resolved channel where such a jet is not found but instead a pair of small-radius ($R=0.4$) jets with invariant mass $65~\text{GeV} < m_{jj} < 110~\text{GeV}$ exists. In the latter case, $V$ is reconstructed from this jet pair (see appendix~\ref{sec:recast} and ref.~\cite{CMS:2016tio} for details). The boosted channel is the one where the CMS excess is most prominent. We give our results in figure~\ref{fig:llJ_CMS}, without separation into $b$-tagged and untagged samples.
Although $WZX$ production alone clearly cannot explain the size of this excess (around 30 events), and additional sources are needed for the production of the $ZV$ resonance, there are several lessons to be drawn:
\begin{itemize}
\item[(i)] The signal is largest in the boosted sample, as it is expected for the decay of a heavy resonance. (We note again that the CMS excess is found mainly in this sample.) The possibility that the CMS excess arises from the decay of a heavy resonance is also in agreement with the fact that no excess was seen in Run 1 analyses.
\item[(ii)] In 57\% of the decays of the heavy resonance $R \to Y^0 \, W \to W Z X$, the particle $X$ is more energetic than the $W$ boson, at the partonic level. Then, if the fat jets they produce are both $V$-tagged, the reconstruction algorithm will select the one resulting from $X$ with a slightly larger probability. This makes, despite the reconstruction inefficiencies, the $\ell \ell J$ peak near the $Y^0$ mass prominent over the continuum at higher $m_{\ell \ell J}$, where the selected jet corresponds to the $W$ boson. In any case, a full detector simulation and accurate energy calibration is necessary to have a more precise prediction of the height of the $Y^0$ peak.
\end{itemize}
If this excess is confirmed, the identity of the particle $X$ and its mass will have to be determined using additional searches in leptonic decay modes.

\subsection{ATLAS Run 2 search in the $\ell\ell J$ channel}

\begin{figure}[htb]
\begin{center}
\begin{tabular}{ccc}
\includegraphics[height=5.2cm,clip=]{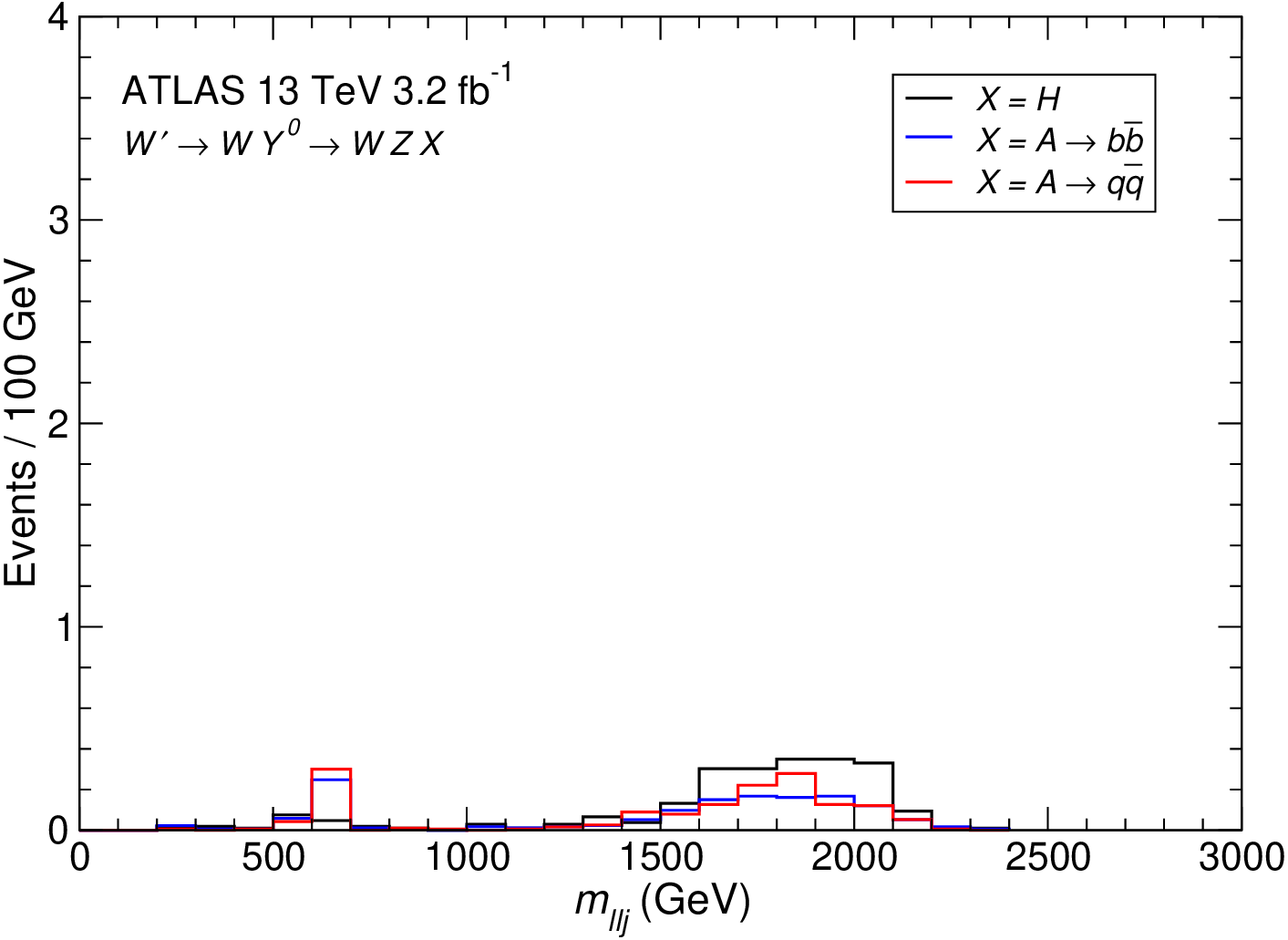} & \quad &
\includegraphics[height=5.2cm,clip=]{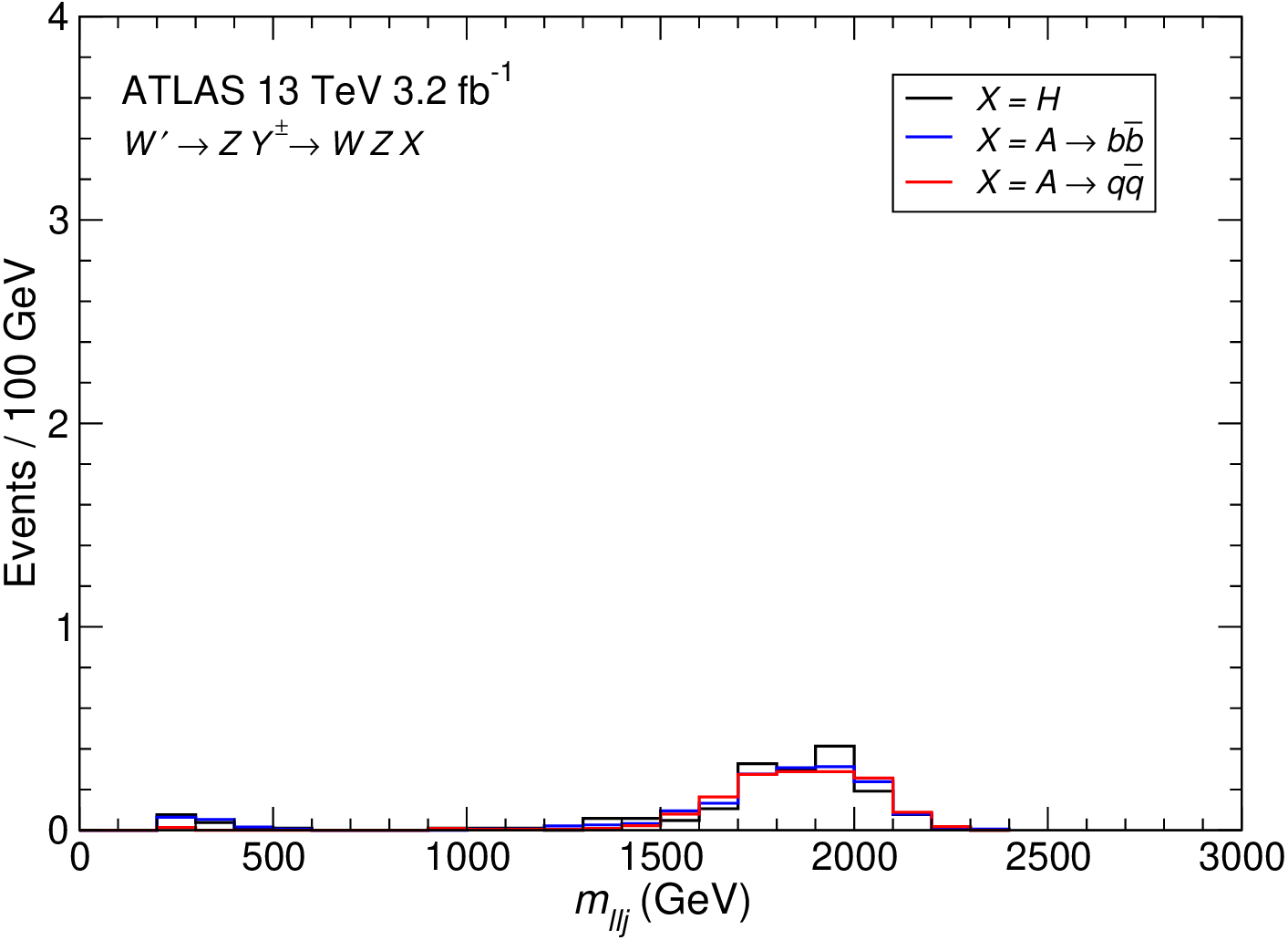}
\end{tabular}
\caption{Reconstructed $\ell \ell J$ mass distribution for triboson signals in the ATLAS Run 2 $\ell \ell J$ analysis~\cite{ATLASllJ2015}, in the $WZ$ selection.}
\label{fig:llJ}
\end{center}
\end{figure}

The ATLAS search for diboson resonances in this final state uses an event selection targetting high masses. Its sensitivity to triboson signals is poor not only because the $Z$ boson leptonic branching ratio is small but also because the $m_{\ell \ell J}$ distribution does not exhibit a peak at the heavy resonance mass, which would be relatively easy to spot due to the smaller background. In addition, the sensitivity is reduced with respect to diboson resonances by the requirement in the event selection that the dilepton pair and the boson-tagged jet have transverse momenta $p_T > 0.4 \, m_{\ell \ell J}$. We present the results of our simulations conveniently normalised in figure~\ref{fig:llJ} for the $WZ$ selection, namely asking that the jet mass lies in a window of $\pm 15$ GeV around the expected $W$ mass peak. As anticipated, the sensitivity is very poor, and the prediction of one event in the full range $1.5-2$ TeV is quite compatible with the null result of ATLAS searches in this channel.

\subsection{ATLAS searches for $VH$ final states}

\begin{figure}[htb]
\begin{center}
\begin{tabular}{ccc}
\includegraphics[height=5.2cm,clip=]{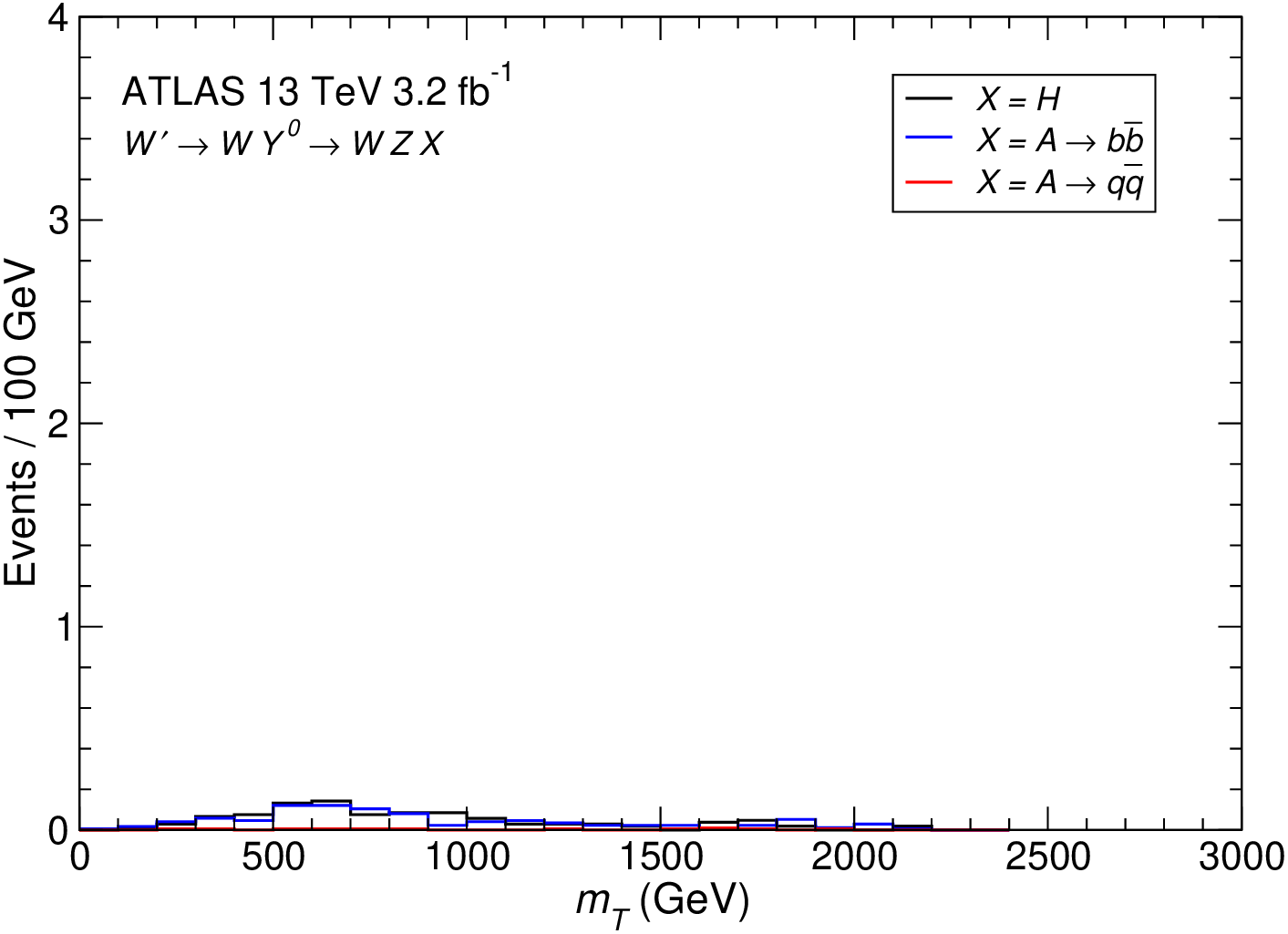} & \quad &
\includegraphics[height=5.2cm,clip=]{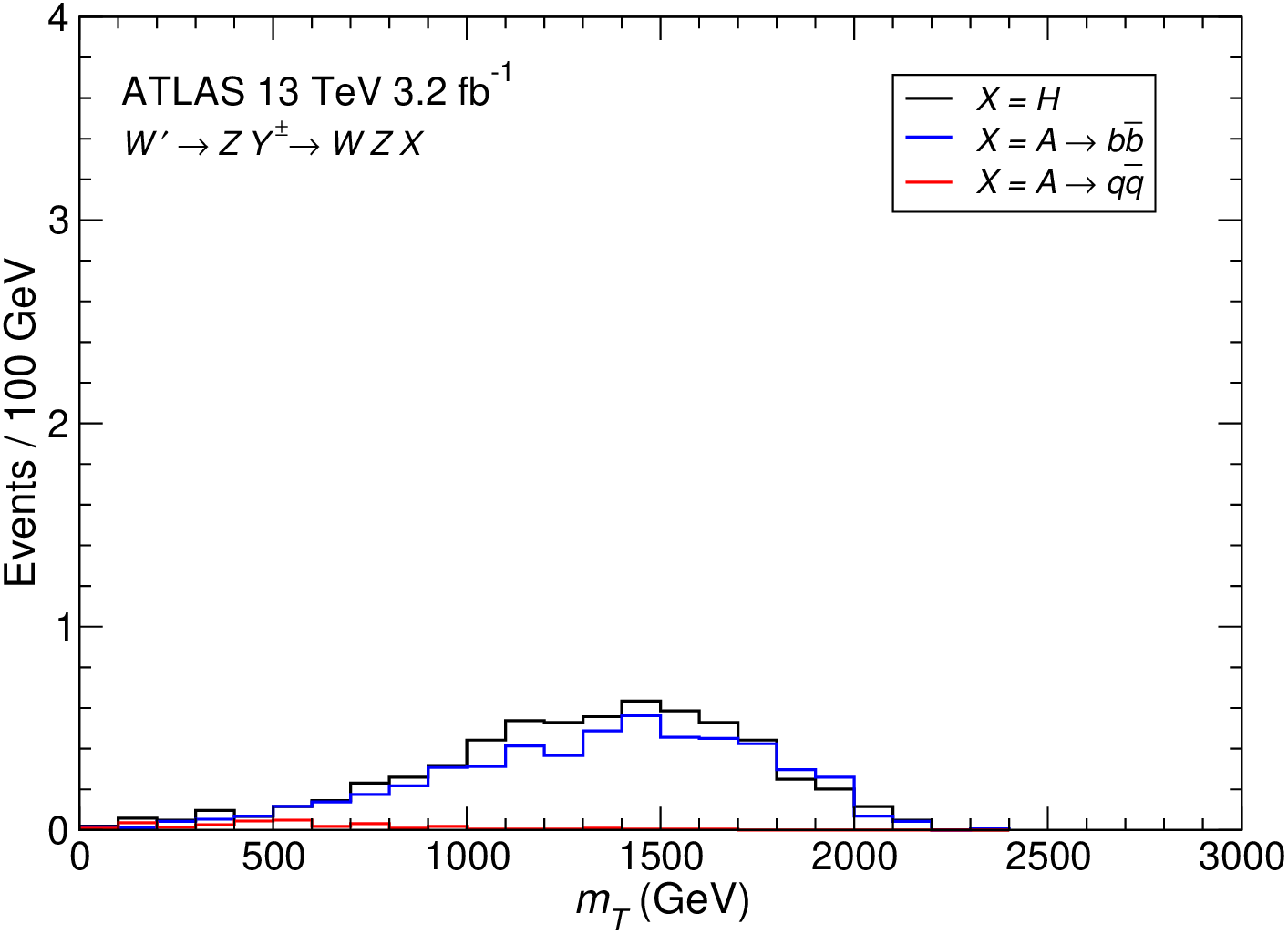} \\
\includegraphics[height=5.2cm,clip=]{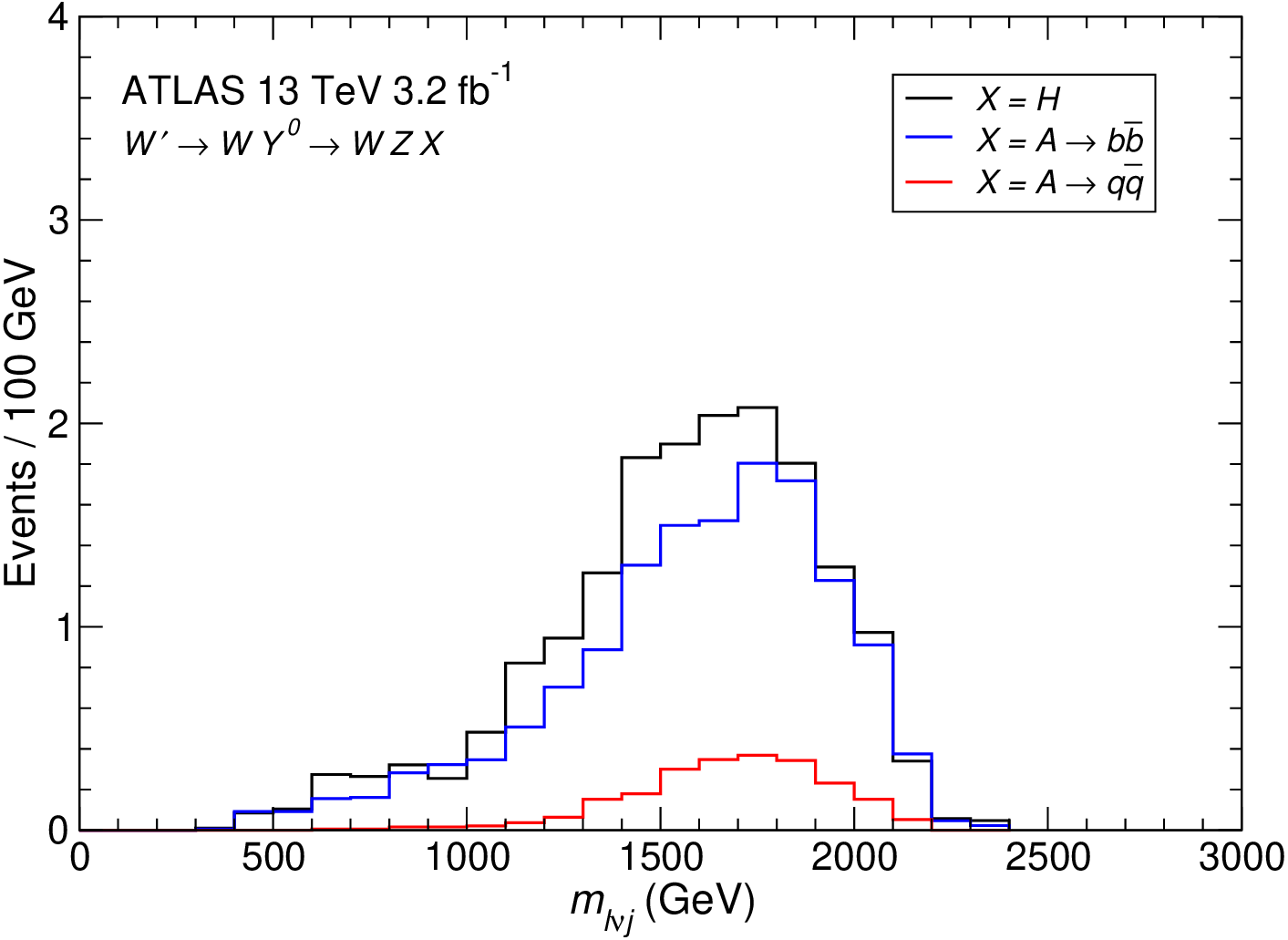} & \quad &
\includegraphics[height=5.2cm,clip=]{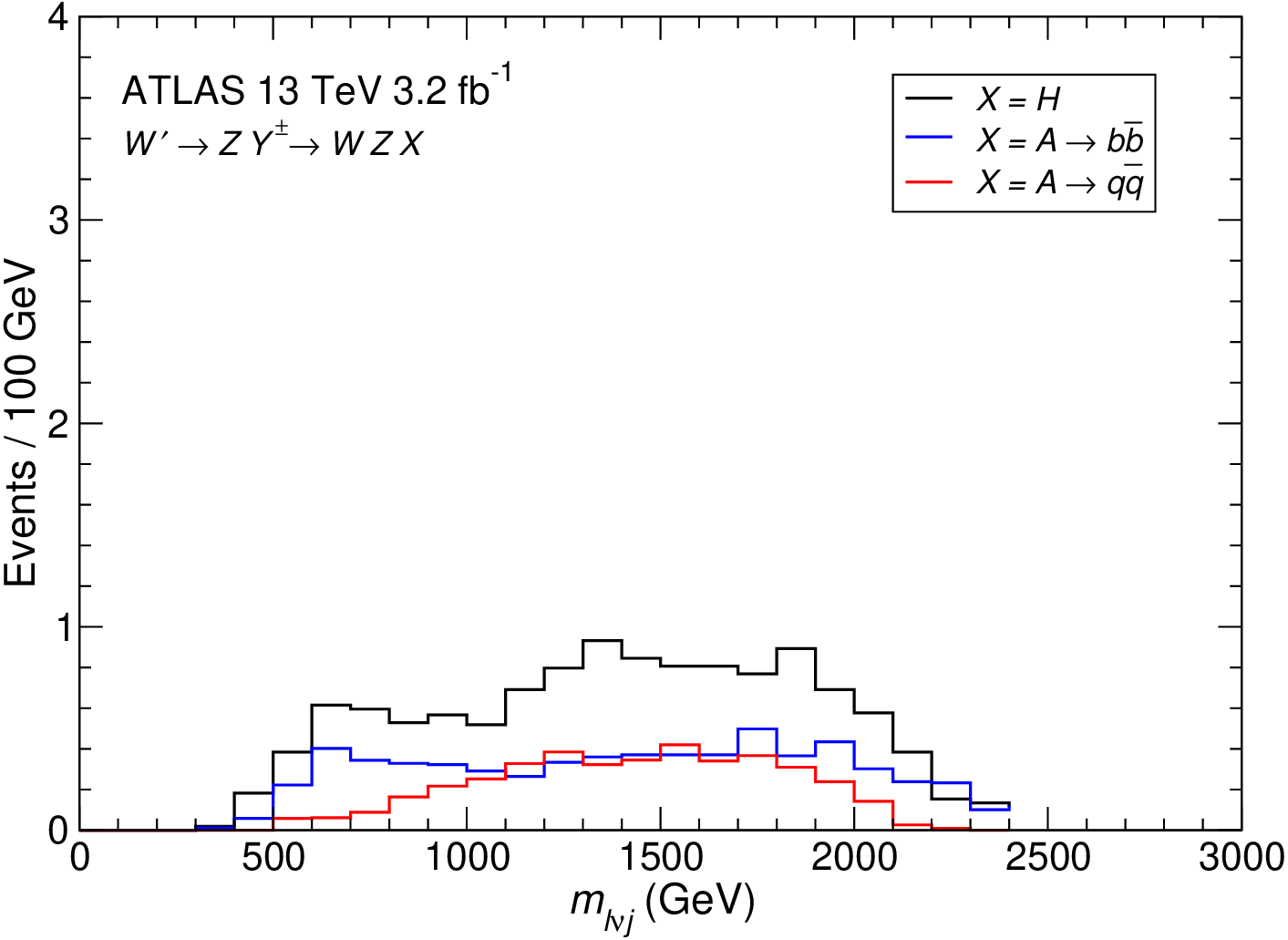} \\
\includegraphics[height=5.2cm,clip=]{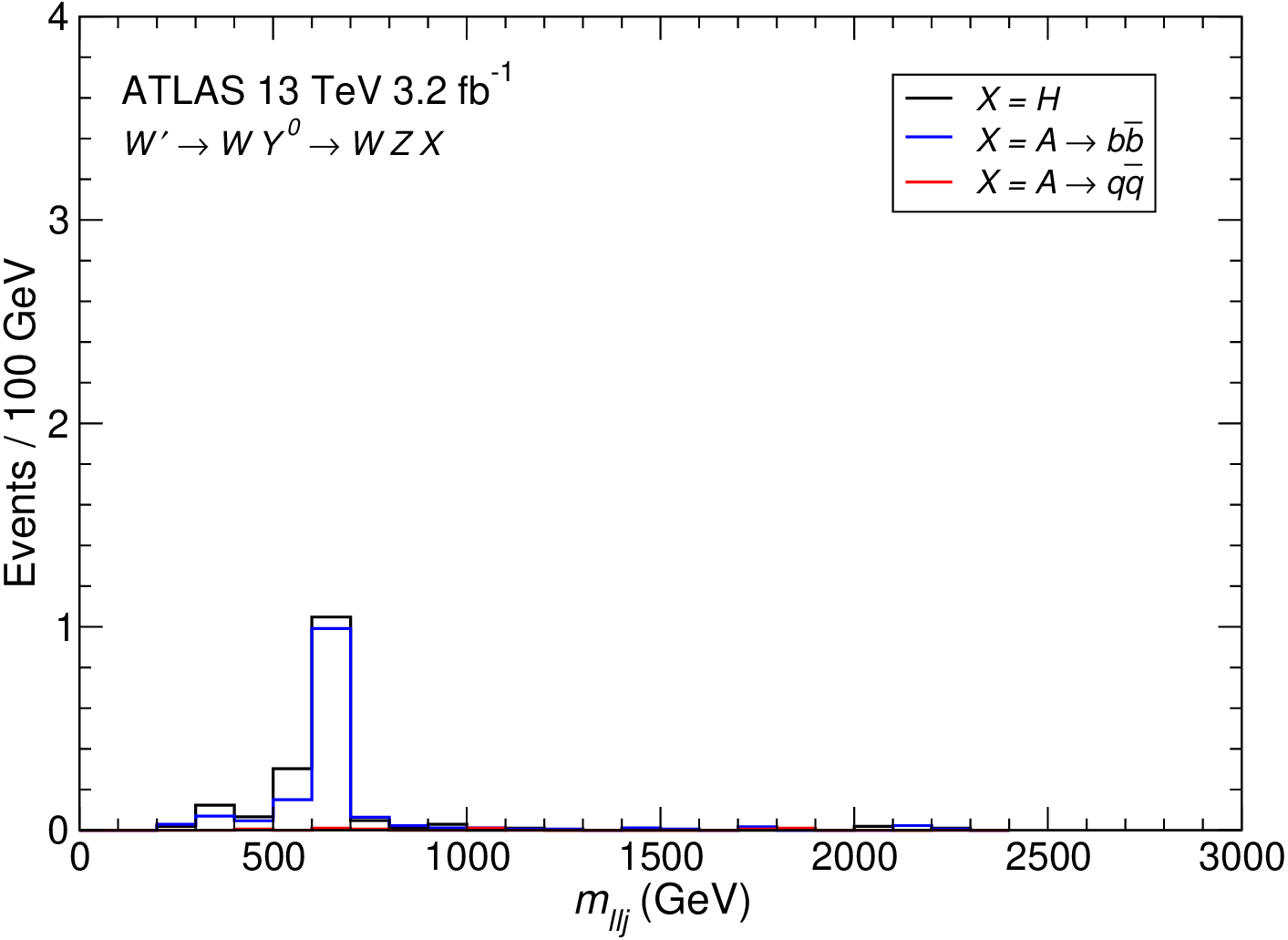} & \quad &
\includegraphics[height=5.2cm,clip=]{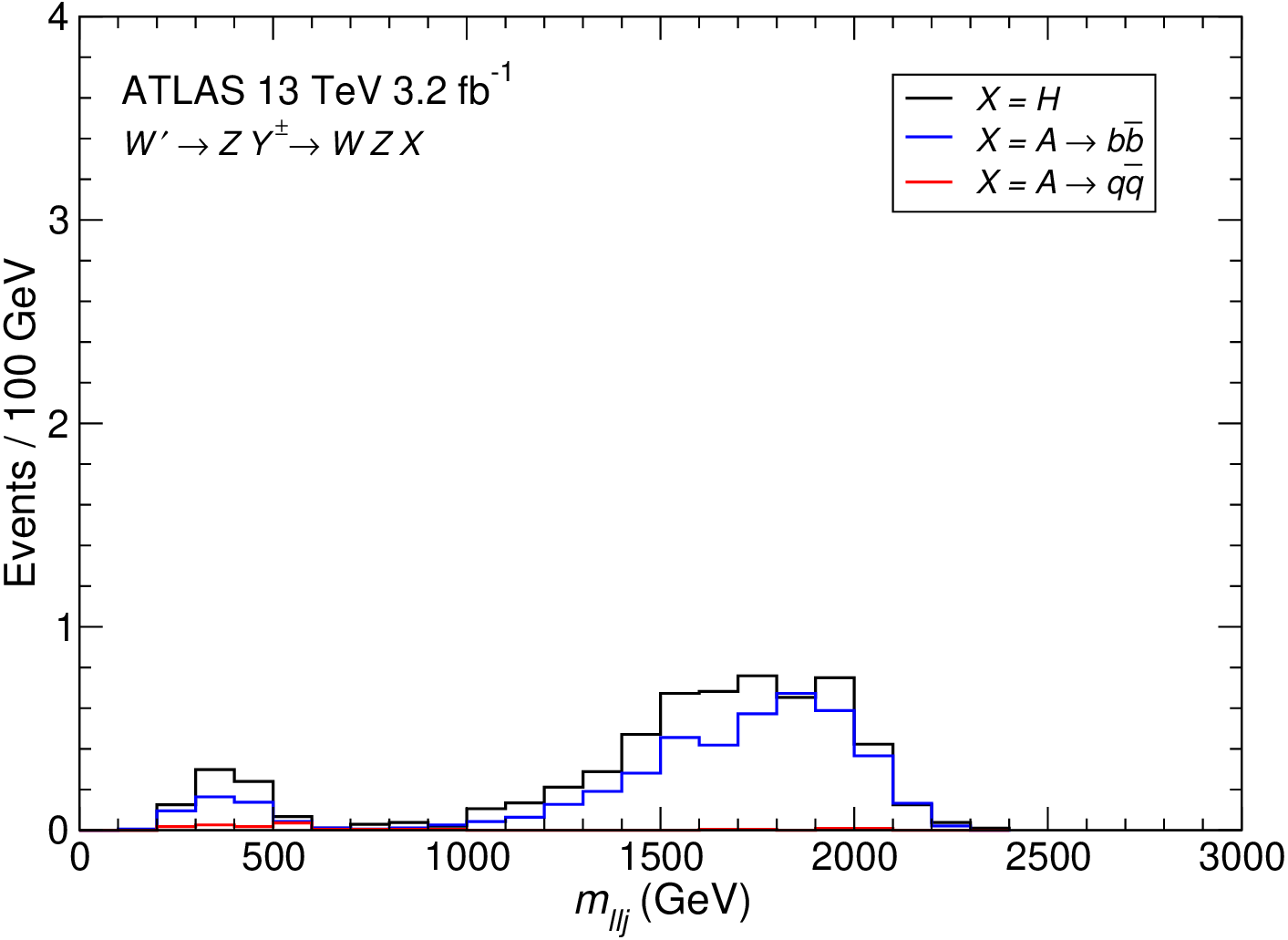}
\end{tabular}
\caption{Reconstructed invariant masses for the ATLAS $VH$ analyses~\cite{Aaboud:2016lwx} Top: transverse mass distribution for the 0-lepton final state. Middle: reconstructed $\ell\nu J$ invariant mass distribution for the 1-lepton final state. Bottom: reconstructed $\ell \ell J$ invariant mass distribution for the 2-lepton final state. }
\label{fig:VH}
\end{center}
\end{figure}

In addition to the searches for a heavy resonance decaying into two gauge bosons, the ATLAS and CMS Collaborations have investigated new heavy resonances decaying into a gauge boson plus a Higgs boson ($VH$). We use the three analyses in ref.~\cite{Aaboud:2016lwx}, for final states with $0$-leptons, $1$-lepton and $2$-leptons, to test the sensitivity to triboson signals. The $0$-lepton, $1$-lepton and $2$-lepton $VH$ searches are similar to the above discussed $\nu \nu J$, $\ell \nu J$ and $\ell \ell J$ diboson searches, respectively, with the main difference that for the fat jet an invariant mass window of $75-145$ GeV is considered, and $b$ tagging is applied. (In the $1$-lepton $VH$ search $b$ tagging is applied instead of the $b$ veto in the $\ell \nu J$ search.) We present the results of our simulations for these analyses, with the previously used normalisation, in figure~\ref{fig:VH}. In all cases the predictions of our benchmark scenarios are compatible with the results of the searches, including the one-lepton channel where a handful of events above the SM prediction are observed in the $1.5-2$ TeV range.\footnote{We have to bear in mind that the efficiency obtained in our simulations for a $WH$ signal in this channel is 50\% larger than the one reported by the ATLAS Collaboration (see appendix~\ref{sec:recast}). Moreover, the signal populates the $W$ and top control regions as for the $\ell \nu J$ channel.} We also note that in the $2$-lepton final state, for $Y^0 \to ZH$ and $Y^0 \to ZX$, $X \to b \bar b$ a small peak is produced at the $Y^0$ mass, although the analysis is not optimised for the low-mass region and this peak is invisible in data.

\section{Triboson discovery strategies}
\label{sec:5}

The identification of a triboson resonance in the current ATLAS and CMS analyses in the fully hadronic channel is not easy, as they search for an excess in the invariant mass distribution of (only) two boson-tagged jets. One strategy that one might consider applying in the ATLAS searches is to remove the $p_T$ asymmetry cut to look for an enhancement at lower dijet invariant masses, as predicted by the simpler analysis performed in ref.~\cite{Aguilar-Saavedra:2015rna}. However, the size of this enhancement is somewhat dependent on the mass of the intermediate particle $Y^0 / Y^\pm$ and the details of the jet filtering procedure, and can only be estimated with a more detailed simulation like the one performed here. This is so because, as mentioned before, the kinematical configurations where two of the bosons merge into a single jet may also pass the event selection criteria, due to the jet filtering performed, and these configurations produce two fat jets with similar transverse momentum.

\begin{figure}[t]
\begin{center}
\begin{tabular}{ccc}
\includegraphics[height=5.2cm,clip=]{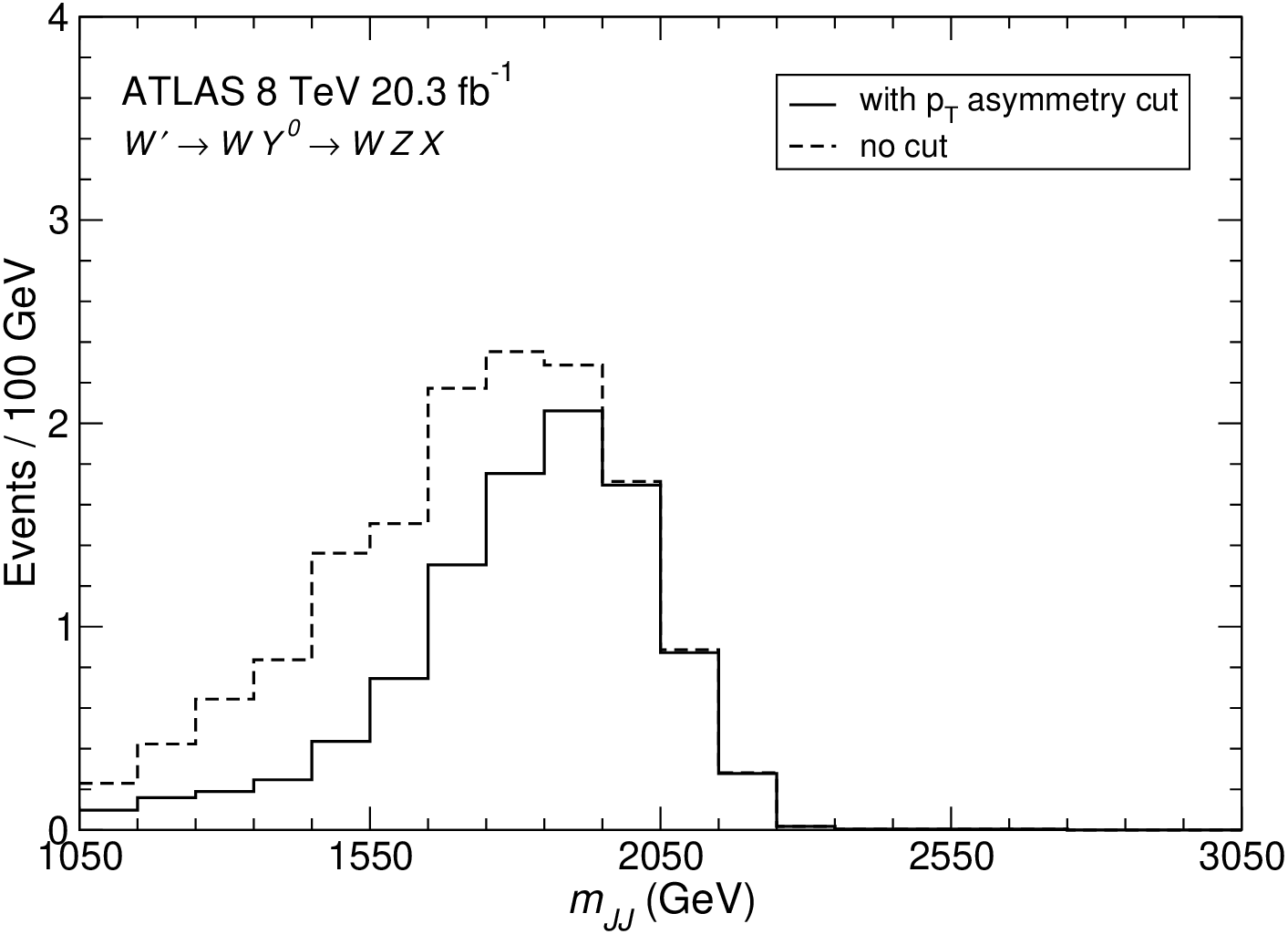} & \quad &
\includegraphics[height=5.2cm,clip=]{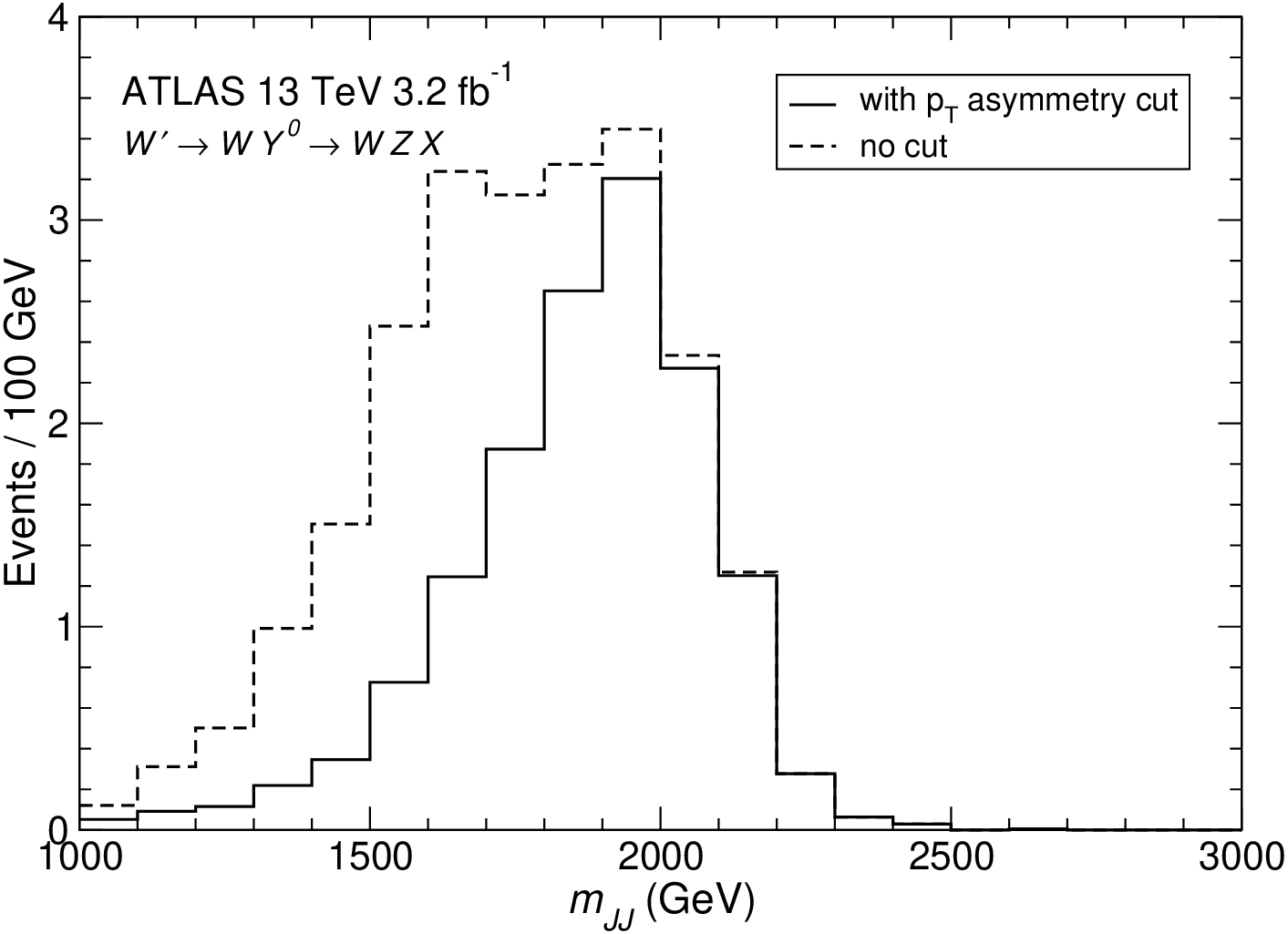}
\end{tabular}
\caption{Dijet invariant mass distribution for a selected triboson signal (see the text) in the ATLAS Run 1 (left) and Run 2 (right) fully hadronic analyses, in the $WZ$ selection without $\ntrk$, with and without the $p_T$ asymmetry cut.}
\label{fig:noPT}
\end{center}
\end{figure}

We give in figure~\ref{fig:noPT} the dijet invariant mass distributions with and without the $p_T$ asymmetry cut for the benchmark with $Y^0 \to ZA$, $A \to b \bar b$, after applying the event selection of the ATLAS Run 1 and Run 2 analyses. We see that the effect of the removal of this requirement is not as pronounced as it was obtained in ref.~\cite{Aguilar-Saavedra:2015rna} with a simpler simulation, especially at Run 1. Moreover, the effect is also mass-dependent, so this test cannot be used to probe the presence of a triboson resonance.

\begin{figure}[t!]
\begin{center}
\includegraphics[height=5.2cm,clip=]{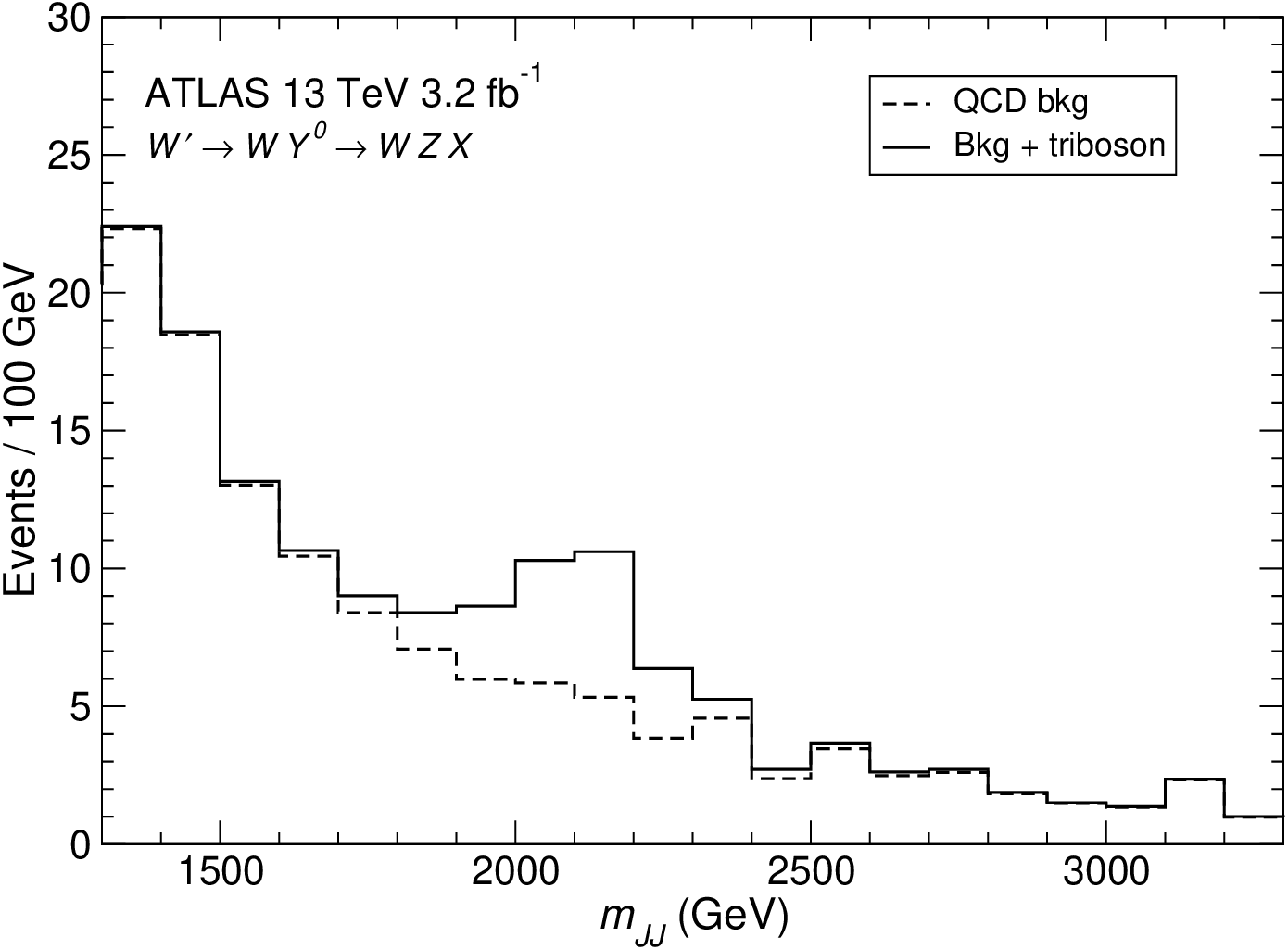} 
\caption{Trijet invariant mass distribution for the SM QCD background and the background plus a triboson signal in the ATLAS Run 2 fully hadronic analysis, in the $WZ$ selection without $\ntrk$ and without the $p_T$ asymmetry cut.}
\label{fig:m3J}
\end{center}
\end{figure}

A very simple modification of the ATLAS Run 2 analysis that would enhance the significance of a triboson signal would be to ask in the event selection for the presence of a third softer jet, say with $p_T \geq 50$ GeV, and consider as discriminating variable the three-jet invariant mass, keeping the rest of the event selection criteria, except the $\ntrk$ cut, and optionally removing the $p_T$ asymmetry cut. We show in figure~\ref{fig:m3J} the results of our simulations with such selection, for the case $Y^0 \to Z A$, $A \to b \bar b$.
It is found that 98\% of the signal events have such a third jet, therefore the excess is magnified and more localised, with 15 extra events at invariant mases between 1.9 and 2.3 TeV. We have performed pseudo-experiments by taking the numbers of expected events per bin as the mean of a Poisson distribution, and have applied the same likelihood ratio as in section~\ref{sec:2} to estimate the signal significance. We find that the expected signal significance is of $4\sigma$ for a luminosity of 15 fb$^{-1}$. A similar strategy can be applied to the CMS $VV$ search in the hadronic channel.

One can also consider dropping one of the cuts on jet masses to enhance the signal. This is advantageous for lower $Y^\pm$, $Y^0$ masses where its decay products often merge into a single jet, but for the benchmark scenarios studied here with $M_R = 2.25$ TeV, $M_{Y^\pm,} = M_{Y^0} = 650$ GeV, the increase in the background is much larger than the signal enhancement. Nevertheless, this modification of the analysis strategy might be convenient to study diboson decays of the heavier $Z'$ boson (see the next section).

\section{Discussion}
\label{sec:6}

In the previous sections we have shown that a triboson signal may explain the excesses observed in the ATLAS Run 1 and Run 2 analyses --- observed provided a cut on $\ntrk$ is not applied in the Run 2 analysis --- while still being compatible with a variety of other diboson and $VH$ searches performed. The reason why the application of an upper cut on $\ntrk$ removes the Run 2 excess, if not a mere statistical effect, remains to be determined. We now turn to the question of the necessary coupling to produce such a signal. As before, we take as reference a signal normalisation of 10 events in the dijet invariant mass interval $1700-2100$ GeV, and study in detail the case $Y^0 \to ZX$, with $X=A \to b \bar b$. In this $m_{JJ}$ interval the signal efficiency for this benchmark scenario is 0.035, relative to all decays of the $WZX$ sample, from which we obtain the necessary cross section
\begin{equation}
\sigma(pp \to R \to WZX) = 90~\text{fb} \,.
\end{equation}
For the $W'$ boson of an extended $\text{SU}(2)_R$ group with gauge coupling $g_R$, the production cross section is $\sigma(pp \to W') \simeq 2.45\,g_R^2~\text{pb}$ at 13 TeV. In a minimal LR extension of the SM, the heavy scalar $H_1^0$ can play the role of the neutral resonance $Y^0$, with a branching ratio $\text{Br}(W' \to H_1^0 \, W) \simeq 0.02$~\cite{Aguilar-Saavedra:2015iew}. Taking this value as reference example, the necessary coupling $g_R \simeq 2 \,g_W$ ($g_W$ being the weak constant) required to reproduce the 2 TeV excess is a bit large, implying in particular a large $W'$ width $\Gamma_{W'} \sim 270~\text{GeV}$.\footnote{Searches for $W' \to t\bar b$ do not pose a problem for a large $g_R$ since the partial width for specific states depends on the right-handed quark mixing matrix.} Nevertheless, if the quark sector comprises new vector-like singlets --- as in grand unified extensions of the SM gauge group --- the coupling of the $W'$ boson to SM quarks and correspondingly the decay width into $q \bar q'$ are reduced, resulting in a smaller $W'$ width and enhanced branching ratio for $W' \to H_1^0 \, W$~\cite{Collins:2015wua}. The mixing with new quarks also allows to loosen constraints from dijet production~\cite{Buen-Abad:2016myz}.

The extension of the SM gauge group with an extra $\text{SU}(2)_R$ also contains a new neutral boson $Z'$. The relation between its mass and the $W'$ mass depends on the particular breaking mechanism of this extra symmetry,
\begin{equation}
M_{Z'} = \frac{k}{\sin \varphi} M_{W'} \,,
\end{equation}
with $k=\sqrt 2\;(1)$ if the breaking of $\text{SU}(2)_R$ is due to a scalar triplet (doublet), and $\sin \varphi$ a mixing angle determined by the gauge couplings and the SM weak mixing angle by
\begin{equation}
\cos \varphi = \frac{g_L}{g_R} \tan \theta_W \simeq 0.55 \, \frac{g_L}{g_R}
\end{equation}
(see for example ref.~\cite{Aguilar-Saavedra:2015iew} for additional details). For the $W'$ mass taken in the benchmark scenarios studied, $M_{W'} = 2.25~\text{TeV}$, and a coupling $g_R \sim 2 g_L $ that could explain the excess, we would then have $\sin \varphi \simeq 1$ and $M_{Z'} \simeq 3.2$ TeV if the $\text{SU}(2)_R$ breaking is mediated by a triplet. A $Z'$ particle of this mass and with the couplings to leptons predicted by the minimal LR model is in conflict with direct searches in the leptonic decay channels; however, these limits would again be relaxed in non-minimal models with a different matter content, or if matter is placed in a different representation of the $\text{SU}(2)_R$ group.

As we have mentioned, the lighter particle $Y^0$ involved in the heavy resonance decay can be the same particle involved in the CMS excess at 650 GeV~\cite{CMS:2016tio}, if for example it decays into $ZA$, with $A \to b \bar b$. (The possibility that the $X$ particle is the $Z$ boson is disfavoured by the results of the $\ell \nu J$ search, since the main decay channel of the $Z$ boson is into light quarks, and also by the absence of multileptonic signals, but a dedicated analysis is necessary to draw any conclusion.) Although in section~\ref{sec:CMSllJ} we have seen that the number of events predicted from the triboson decay of the 2.25 TeV resonance is far too small to account for it, other sources of $Y^0$ particles may also contribute, since the search is inclusive. These additional sources would comprise additional decay modes of the 2.25 TeV resonance as well as of heavier ones. In this regard, it is worthwhile pointing out that the $Z'$ decays into SM and heavier bosons\footnote{For example, $Z' \to W^+ W^-$, $Z' \to ZH $, $Z' \to Z H_1^0$, $Z' \to H A^0$, $Z' \to H_1^0 A^0$, $Z' \to H^+ H^-$ in the specific context of a LR model.}
would also be visible in the hadronic diboson searches provided a different analysis strategy is adopted, removing for example one of the cuts on jet masses as mentioned in section~\ref{sec:5}.

To conclude, in this paper we have shown via detailed simulations the capability for the triboson scenarios to explain a few excesses found in hadronic diboson searches while remaining consistent with null results in other channels. In order to determine whether such excesses indeed correspond to new physics and to discover their physical origin, it is desirable for experiments to consider triboson search strategies in addition to traditional diboson benchmarks. We have investigated a simple modification of the ATLAS hadronic search that makes it more sensitive to this type of resonance. The same strategy could be applied for the CMS search as well. In most of the remaining diboson searches the analysis modifications are obvious, having in mind that a triboson resonance yields not only the two bosons searched for, but also an additional particle.

{\bf Note added. } After the submission of this paper, updated experimental results were released with the first 2016 data for the ICHEP conference in August, with integrated luminosities $4-5$ times larger than those discussed in this paper. Still, the status of diboson searches has not dramatically changed. In the ATLAS fully hadronic search, the local significance near 2 TeV with the nominal $WZ$ selection has grown from around $1\sigma$ with 3.2 fb$^{-1}$ to $1.9\sigma$  with 15.5 fb$^{-1}$. But results without the $\ntrk$ cut, which would confirm or disprove some of the conjectures made in this paper, have not been released by the ATLAS Collaboration. The CMS Collaboration has not yet provided updated results for the fully hadronic channel either. Among the remaining final states, the most interesting one is a new $VH$ search in the hadronic channel~\cite{ATLAS:2016kxc} discussed in detail in appendix~\ref{sec:add}.
In the rest of analyses the absence of excesses is compatible with the presence of a triboson resonance despite the larger luminosities, since (i) the predicted numbers of events are generally small, and (ii) the SM backgrounds are frequently normalised from data in phase space regions where a triboson resonance has a non-negligible contribution.

\section*{Acknowledgelements}We thank M. Perelstein and S. Santos for helpful discussions. This work has been supported by MINECO (Spain) project FPA2013-47836-C3-2-P (including ERDF), and by Junta de Andaluc\'{\i}a project FQM 101. JC and SL are supported in part by the NSF through grant PHY-1316222.

\appendix
\section{Recast of existing diboson searches}
\label{sec:recast}

Here we provide further details on the recasts of the diboson searches performed in this paper, which include Run 1 and Run 2 hadronic diboson searches from both the ATLAS~\cite{Aad:2015owa,ATLASJJ2015} and CMS~\cite{Khachatryan:2014hpa,CMS:2015nmz} Collaborations. Additionally, we reproduce the Run 2 searches in the $\nu \nu J$~\cite{ATLASvvJ2015}, $l \nu J$~\cite{ATLASlvJ2015}, $l l J$~\cite{ATLASllJ2015,CMS:2016tio}, and $VH$~\cite{Aaboud:2016lwx} channels. We use different {\scshape Delphes} detector cards for each search to incorporate the correct charged lepton isolation requirements.

\subsection{ATLAS hadronic diboson channels}

We here describe the Run 2 search for brevity, and refer the reader to ref.~\cite{Aad:2015owa} for the details of the very similar Run 1 search. For this analysis, ATLAS employs large-$R$ jets (anti-$k_T$, $R = 1.0$), denoted as $J$. Large-$R$ jets are trimmed by re-clustering the large-$R$ jet constituents using the anti-$k_T$ algorithm with $R=0.2$ and dropping any of the sub-jets with $p_T$ less than 5\% of the original jet $p_T$. Boson candidate jets are tagged for two-prong substructure by imposing an upper cut on the $D_2^{(\beta=1)}$ variable, abbreviated here as $D_2$. This cut is $p_T$-dependent~\cite{ATLASD2}, and is approximately $D_2 < 1 + 0.8 \, (p_T - 300) /1200$, with $p_T$ in GeV. Events with charged leptons or $E_{T}^{\text{miss}} > 250$ GeV are vetoed. 

Events must contain two jets $J$ with $p_T > 200$ GeV and $|\eta| < 2.0$. Furthermore, it is required that the the leading large-$R$ jet has $p_T > 450$ GeV and the leading and sub-leading large-$R$ jets $J_1$, $J_2$ have
\begin{itemize}
	\item invariant mass $m_{JJ} > 1$ TeV;
	\item small rapidity separation $|\Delta y_{12}| < 1.2$;
	\item transverse momentum asymmetry $\displaystyle \frac{p_{T1}-p_{T2}}{p_{T1}+p_{T2}} < 0.15$.
\end{itemize}
The jets are tagged as $W$ or $Z$ candidates if their mass is within an interval of $\pm 15$ GeV around the expected resonance peak. Notice that a jet can be simultaneously tagged as $W$ and $Z$.
For a $W' \to WZ$ diboson signal with $M_{W'} = 2$ TeV, and considering only $W$ and $Z$ hadronic decays, we obtain efficiencies of 0.16 and 0.13 without and with the cut on $\ntrk$. These efficiencies are slightly smaller than the ones obtained by the ATLAS Collaboration for this type of signal, 0.2 and 0.15, respectively~\cite{ATLASJJ2015}.

%%%%%%%%%%%%%%%%%%%%%%%%%%%%%%%%%%%%%%%

\subsection{CMS hadronic diboson channel}

The Run 2 CMS hadronic search uses large-$R$ anti-$k_T$ jets with $R = 0.8$ (referred to as AK8 jets) for hadronic vector boson candidates. Events containing an isolated charged lepton are rejected. Charged lepton isolation requirements are based on reconstructed tracks and calorimeter deposits within $\Delta R = 0.3$. A jet-pruning algorithm is performed on the AK8 jets, which starts from all original constituents of the jet and discards soft recombinations after each step of the Cambridge-Aachen algorithm. Only jets with $|\eta| < 2.4$ are considered.

$N$-subjettiness is used to classify jet substructure, in particular $\tau_{21} = \tau_2 / \tau_1$. AK8 jets with $\tau_{21} > 0.75$ are rejected. 
The high-purity (HP) category of AK8 jets is defined by $\tau_{21} < 0.45$, and the low-purity (LP) category by $0.45 < \tau_{21} < 0.75$. A jet is considered as tagged if it has $\tau_{21} < 0.75$ and mass in the range $65~\text{GeV} < m_J < 105~\text{GeV}$, and it has momentum $p_T > 200$ GeV. A jet is considered as $W$-tagged if its mass is in the range $65-85$ GeV, and $Z$-tagged if it is in the range $85-105$ GeV. 

For this analysis, a HP sample is defined by the presence of two boson jets being tagged as HP, whereas a LP sample is defined where one of them is of HP and the other one LP. The leading two jets within these categories must have
\begin{itemize}
\item rapidity difference $|\Delta y_{12}| \leq 1.3$;
\item dijet invariant mass $m_{JJ} \geq 1$ TeV.
\end{itemize}
The efficiency found for a 2 TeV $W'$ boson decaying into $WZ$, including all decays of the $W$ and $Z$ bosons, is of 0.040, quite close to the efficiency of 0.038 obtained by the CMS Collaboration in ref.~\cite{CMS:2015nmz}.

The CMS Run 1 analysis is similar, with the main differences that (i) the leading and sub-leading AK8 jets are considered and a cut on $\tau_{21}$ is imposed to select HP (or LP) samples, in contrast with the Run 2 strategy of considering all $V$-tagged AK8 jets and selecting the leading and sub-leading one among them; (ii) a single mass window $70-100$ GeV is considered and $W/Z$ discrimination is not attempted.

%%%%%%%%%%%%%%%%%%%%%%%%%%%%%%%%%%%%%%%

\subsection{ATLAS $\nu \nu J$ diboson channel}

For this search, we remove the lepton isolation requirements within the {\scshape Delphes} ATLAS detector card in order to best match the loose track-based isolation criteria used in the analysis. In this channel, events with charged leptons are vetoed. There are two jet definitions used in this analysis: large-$R$ jets (anti-$k_T$, $R = 1.0$) and small-$R$ jets (anti-$k_T$, $R=0.4$). The leading large-$R$ jet is trimmed as in the ATLAS fully hadronic channel and, after trimming, it is required to have $p_T > 200$ GeV and $|\eta| < 2$. The leading jet is also tagged for two-prong substructure by imposing an upper cut on $D_2$ and its mass must be consistent with the $W$ or $Z$ boson mass within a window of $\pm 15$ GeV.

The missing energy $E_{T}^{\text{miss}}$ is calculated from the vectorial sum of charged leptons and small-$R$ jets, and event selection also requires
\begin{itemize}
	\item large $E_{T}^{\text{miss}} > 250$ GeV;
	\item $E_{T}^{\text{miss}}$ isolation: the minimum azimuthal separation between the $E_{T}^{\text{miss}}$ vector and any small-$R$ jet is required to be greater than 0.6 radians.
\end{itemize}
The discriminant variable in this channel is the transverse mass calculated as
\begin{equation}
m_T = \sqrt{ (E_{T,J} + E_{T}^{\text{miss}} )^2 - ( \vec{p}_{T,J})^2 + (\vec{E}_{T}^{\text{miss}})^2} \,, \end{equation}
where $E_{T,J} = \sqrt{ m_J^2 + p^2_{T,J} }$. With this event selection we obtain for a $W' \to WZ$ diboson signal with $M_{W'} = 2$ TeV an efficiency of 0.083, relative to the sample with all $W$ and $Z$ decays. This value is slightly larger than the efficiency of 0.05 found by the ATLAS Collaboration for this type of signal~\cite{Aaboud:2016okv}.

%%%%%%%%%%%%%%%%%%%%%%%%%%%%%%%%%%%%%%%

\subsection{ATLAS $\ell \nu J$ diboson channel}

This search again employs the large-$R$ and small-$R$ jet definitions as above. 
The leading large-$R$ jet is trimmed and, after trimming, it is required to have $p_T > 250$ GeV and $|\eta| < 2$, and tagged for two-prong substructure by imposing an upper cut on $D_2$. Its mass must be consistent with the $W$ or $Z$ boson mass within a window of $\pm 13$ GeV. Small-$R$ jets are used for $b$-tagging, and events with $b$-tagged jets are rejected. Leptons must be isolated from tracks with $p_T > 1$ GeV and calorimeter activity within $\Delta R = 0.2$ of the lepton. Additional selection criteria are
\begin{itemize}
\item One charged lepton with $p_T > 25$ GeV and $|\eta| < 2.5$ for muons, $|\eta| < 1.37$ or $1.52 < |\eta| < 2.47$ for electrons;
	\item large $E_{T}^{\text{miss}} > 100$ GeV;
	\item transverse momentum $p_T > 200$ GeV and $p_T > 0.4\, m_{\ell \nu J}$ for the $\ell \nu$ pair, with $m_{\ell \nu J}$ the $\ell \nu J$ reconstructed mass;
	\item transverse momentum $p_T > 0.4\, m_{\ell \nu J}$ for the leading jet as well.
	\end{itemize}
The $\ell \nu J$ reconstructed mass $m_{\ell \nu J}$ introduced above is computed imposing that the $\ell \nu$ pair results from a $W$ boson decay, solving a quadratic equation for the neutrino longitudinal momentum and taking the solution with minimum neutrino longitudinal momentum if both are real, or the real part if they are complex. For this channel we obtain for a $W' \to WZ$ diboson signal with $M_{W'} = 2$ TeV an efficiency of 0.053, relative to the sample with all $W$ and $Z$ decays, close to the ATLAS reported efficiency of around 0.06~\cite{Aaboud:2016okv} for a $W'$ boson signal of this mass.

%%%%%%%%%%%%%%%%%%%%%%%%%%%%%%%%%%%%%%%

\subsection{ATLAS $\ell \ell J$ diboson channel}

For this search, we remove the lepton isolation requirements within the {\scshape Delphes} ATLAS detector card in order to best match the loose isolation criteria used in the analysis. The same large-$R$ jets with $|\eta| < 2.0$ are used as in other ATLAS analyses. The leading large-$R$ jet must have $p_T > 200$ GeV and two-prong structure (achieved by a cut on the $D_2$ variable), and a mass consistent with the $W$ or $Z$ boson mass within a window of $\pm 15$ GeV. The remaining selection criteria are:
\begin{itemize}
\item two same-flavour opposite-sign leptons with $p_T > 25$ GeV and $|\eta| < 2.5$ for muons; $|\eta| < 1.37$ or $1.52 < |\eta| < 2.47$ for electrons;
\item dilepton invariant mass $66~\text{GeV} < m_{\mu \mu} < 116~\text{GeV}$ for muons, and $83~\text{GeV} < m_{ee} < 99~\text{GeV}$ for electrons;
\item for the leading jet and the $\ell \ell$ pair $p_T > 0.4\, m_{\ell \ell J}$ is required, with $m_{\ell \ell J}$ the $\ell \ell J$ invariant mass.
\end{itemize}
The $\ell \ell J$ invariant mass is simply calculated from the four-momenta of the two charged leptons and the leading fat jet, scaling the lepton momenta so that their invariant mass coincides with the $Z$ boson mass. The efficiency for a $W' \to WZ$ diboson signal with $M_{W'} = 2$ TeV is of 0.01, relative to the sample with all $W$ and $Z$ decays. This number coincides with the ATLAS reported efficiency for this signal~\cite{Aaboud:2016okv}.

%%%%%%%%%%%%%%%%%%%%%%%%%%%%%%%%%%%%%%%

\subsection{CMS $\ell \ell J$ diboson channel}

In the low-mass analysis we focus in, two classes of jets are used: $R=0.8$ jets (AK8 jets) and $R=0.4$ jets (AK4 jets). The high-purity category of AK8 jets is defined by $\tau_{21} < 0.45$. A jet is considered as tagged if it is in the high purity category, its mass is in the range $65~\text{GeV} < m_J < 105~\text{GeV}$, and it has momentum $p_T > 200$ GeV. If such a jet is not found, the reconstruction of the hadronically decaying boson is attempted using pairs of AK4 jets with $p_T > 30$ GeV each, and $p_T(jj) > 100$ GeV, with invariant mass $65~\text{GeV} < m_{jj} < 110~\text{GeV}$. For the charged leptons, the following selection criteria apply:
\begin{itemize}
	\item muons with $|\eta| < 2.4$ and electrons with $|\eta| < 1.44$ or $1.57 < |\eta| < 2.1$ are considered; they must have same flavour and opposite sign, with the leading and sub-leading lepton having with $p_T> 40$ GeV and 24 GeV, respectively;
	\item for electrons, a minimum separation $\Delta R(ee) \geq 0.3$; 
	\item dilepton invariant mass $76~\text{GeV} < m_{\ell \ell} < 106~\text{GeV}$.
\end{itemize}

%%%%%%%%%%%%%%%%%%%%%%%%%%%%%%%%%%%%%%%

\subsection{ATLAS $VH$ searches}

For these searches, three types of jets are employed: large-$R$ jets with $R=1.0$, small-$R$ jets with $R=0.4$ and track jets built from inner detector tracks with $R=0.2$. The three channels require the presence of at least one large-$R$ jet with $p_T > 250$ GeV and $|\eta| < 2.0$. The leading large-$R$ jet is considered as the $H$ candidate, and its mass is required to be $75~\text{GeV} < m_{J} < 145~\text{GeV}$. $b$ tagging is performed by requiring that at least one of its associated track jets is $b$-tagged.

In the 0-lepton channel it is required that no loose leptons are present. For this and the two-lepton channel we remove the lepton isolation requirements within the {\scshape Delphes} ATLAS detector card. Additional requirements are imposed:
\begin{itemize}
\item missing energy $E_{T}^{\text{miss}} > 200$ GeV;
\item $E_{T}^{\text{miss}}$ isolation: the minimum azimuthal separation between the $E_{T}^{\text{miss}}$ vector and the leading large-$R$ jet is required to be greater than $2\pi/3$ radians, and the separation with any small-$R$ jet it is required to be greater than $\pi/9$;
\item events where there is a $b$-tagged track jet not associated to the Higgs candidate are vetoed.
\end{itemize}
The transverse mass is reconstructed as for the $\nu \nu J$ channel. For this channel, the efficiency found for a $Z' \to Z H$ with $M_{Z'} = 1.5$ TeV is 0.045, including all decays of the $Z$ and Higgs boson, comparable to the value of 0.067 obtained by the ATLAS Collaboration~\cite{Aaboud:2016lwx}.

In the 1-lepton channel the presence of exactly one charged lepton with $p_T > 25$ GeV is required, plus $E_{T}^{\text{miss}} > 100$ GeV. Events where there is a $b$-tagged track jet not associated to the Higgs candidate are vetoed. The $\ell \nu H$ mass is reconstructed in an analogous way to the $\ell \nu J$ channel. The efficiency found for a $W' \to W H$ signal with $M_{W'} = 1.5$ TeV is 0.084, slightly larger than the value of 0.051 reported by the ATLAS Collaboration~\cite{Aaboud:2016lwx}.

Finally, in the 2-lepton channel the presence of two same-flavour opposite-charge is required, both with $p_T > 25$ GeV. For this channel we also remove the lepton isolation requirements within the {\scshape Delphes} ATLAS detector card. The invariant mass windows are $70~\text{GeV} < m_{ee} < 110~\text{GeV}$ for electrons and $55~\text{GeV} < m_{\mu \mu} < 125~\text{GeV}$ for muons. The efficiency found for this channel is 0.022 for $Z' \to Z H$ with $M_{Z'} = 1.5$ TeV, slightly larger than 0.014 as obtained by the ATLAS Collaboration~\cite{Aaboud:2016lwx}.

\section{$WZX$ signals and $N_\text{trk}$}
\label{sec:a}

It is conceivable that for a signal consisting of a diboson pair plus extra particles, the different fat jet reconstruction algorithms and boson tagging requirements used by the ATLAS Collaboration in the analysis of Run 1 and Run 2 data might result in a somewhat different efficiency of the $\ntrk$ cut. In order to investigate this possibility, we have considered three representative $WZX$ signals, with (a) $M_X = 100$ GeV, $X \to u \bar u$; (b) $M_X = 100$ GeV, $X \to b \bar b$; (c) $M_X = 300$ GeV, $X \to W^+ W^-$. We have divided the phase space of the $R \to WZX$ decay into 18 regions according to the separation between the three particles $W$, $Z$, $X$ and the energy of $X$. We have considered:
\begin{enumerate}
\item $\Delta \phi_{VV} \equiv \Delta \phi(W,Z)$, the (azimuthal) angle in transverse plane between the $W$ and $Z$ momenta in the laboratory frame. We have separated three regions: $\Delta \phi_{VV} > 2.5$; $2.5 > \Delta \phi_{VV} > 1.2$; (C) $\Delta \phi_{VV} < 1.2$, respectively labelled as `A', `B' and `C'. 
\item $\Delta R_X \equiv \text{min}(\Delta R(X,W),\Delta R(X,Z))$, separating three regions: $\Delta R_X > 1.2$, $1.2 > \Delta R_X > 0.6$; $\Delta R_X < 0.6$, respectively labelled as `1', `2' and `3'.
\item The energy of $X$ in the laboratory frame $E_X$, separating $E_X < 2 M_X$ and $E_X > 2 M_X$.
\end{enumerate}
Among the 18 possible combinations only 10 are relevant; the rest are either kinematically forbidden or the phase space is extremely small.

We have generated $pp \to R \to WZX$ with {\scshape Protos} using a flat matrix element for the decay $R \to WZX$, so as to avoid any bias introduced by the presence of a secondary resonance $Y$. Samples of $5 \times 10^4$ events are generated for each signal and kinematical configuration, for CM energies of 8 and 13 TeV. On the simulated events we have applied all topological cuts and the jet substructure cuts on $\sqrt y$ for ATLAS Run 1 analysis and $D_2$ for ATLAS Run 2 analysis, as well as the jet mass cuts. For events passing these selection criteria, we have evaluated the efficiency $\epsilon_{\ntrk}$ of the $\ntrk$ cut. The results are shown in Tables~\ref{tab:PS1}--\ref{tab:PS3}.

\begin{table}[htb]
\begin{center}
\begin{tabular}{cccccc}
& A1 & A2 & A3 & B1 & B2 \\
$E_X < 2 M_X$ & 0.83 / 0.83 & 0.70 / 0.75 & 0.57 / 0.64 & $\ast$ / $\ast$ & $\ast$ / $\ast$ \\
$E_X > 2 M_X$ & 0.88 / 0.95 & 0.8 / 0.7 & 0.7 / 0.8 & 0.85 / 0.87 & 0.84 / 0.85 \\
\end{tabular}
\end{center}
\caption{Efficiency of the $\ntrk$ cut for events passing the remaining selection criteria of the ATLAS Run 1 / Run 2 analyses, for $WZX$ with $X \to u \bar u$, in several phase space regions described in the text. The asterisks indicate those space regions where the overall efficiency of the remaining selection criteria is smaller than $3 \times 10^{-4}$. \label{tab:PS1}}
\end{table}

\begin{table}[htb]
\begin{center}
\begin{tabular}{cccccc}
& A1 & A2 & A3 & B1 & B2 \\
$E_X < 2 M_X$ & 0.84 / 0.86 & 0.77 / 0.76 & 0.56 / 0.66 & $\ast$ / $\ast$ & $\ast$ / $\ast$ \\
$E_X > 2 M_X$ & 0.89 / 0.88 & 0.7 / 0.8 & 0.7 / 0.8 & 0.89 / 0.88 & 0.87 / 0.89 \\
\end{tabular}
\end{center}
\caption{Same as table~\ref{tab:PS1}, for $WZX$ with $X \to b \bar b$. \label{tab:PS2}}
\end{table}

\begin{table}[htb]
\begin{center}
\begin{tabular}{cccccc}
& A1 & A2 & A3 & B1 & B2 \\
$E_X < 2 M_X$ & 0.89 / 0.87 & 0.85 / 0.83 & 0.8 / 0.7 & 0.9 / 0.9 & 0.9 / $\ast$ \\
$E_X > 2 M_X$ & 0.90 / 0.89 & 0.7 / 0.8 & 0.7 / 0.6 & 0.84 / 0.82 & 0.9 / 0.9 \\
\end{tabular}
\end{center}
\caption{Same as table~\ref{tab:PS1}, for $X \to W^+ W^-$. \label{tab:PS3}}
\end{table}

From this exercise it can be observed that $\epsilon_{\ntrk}$ is very similar for Run 1 and Run 2 selections in most cases, and we do not find any phase space region where there is a significant efficfency drop in the Run 2 selection compared to Run 1. Notice that for those regions where we quote only one significant digit for $\epsilon_{\ntrk}$, the efficiency of the rest of the cuts is rather small, of the order of $10^{-3}$, leading to an uncertainty around $\pm 0.1$ in our evaluation of $\epsilon_{\ntrk}$.

\section{Addendum: ATLAS search for $VH$ production in the fully hadronic channel}
\label{sec:add}

After the submission of this paper, the ATLAS Collaboration released the results of the first search for $VH$ production in the fully hadronic channel~\cite{ATLAS:2016kxc}, which provides one further important test of the triboson hypothesis proposed to explain the $JJ$ excess. In this appendix we give the results of the six triboson scenarios considered for the $VH$ hadronic final state.

The methodology of this search is analogous to other ATLAS diboson and $VH$ analyses. Events must contain two $R=1.0$ jets with $|\eta| < 2.0$, with transverse momentum larger than 450 and 250~GeV, respectively, and no leptons. Among these jets, the one with larger invariant mass is the $H$ candidate whereas the other one is the $V$ candidate. The $V$ candidate must have two-prong structure (achieved by a cut on the $D_2$ variable), and a mass consistent with the $W$ or $Z$ boson mass within a window of $\pm 15$ GeV. For the $H$ candidate, the jet mass is required to be $75~\text{GeV} < m_{J} < 145~\text{GeV}$, and $b$ tagging is performed by requiring that at least one of its associated $R=0.2$ track jets is $b$-tagged. Additional selection criteria are:
\begin{itemize}
\item rapidity difference $|\Delta y_{12}| \leq 1.6$ between the two leading large-$R$ jets;
\item dijet invariant mass $m_{JJ} \geq 1$ TeV;
\item events are rejected if they have $E_{T}^{\text{miss}} > 150$ GeV and the azimuthal separation between the $E_{T}^{\text{miss}}$ vector and the $H$ candidate is smaller than $2\pi/3$ radians.
\end{itemize}
Note that in contrast with the ATLAS search in the fully hadronic mode, a $p_T$ asymmetry cut --- which produces some background shaping --- is not applied here, nor an upper cut on the number of tracks. The efficiency found in our analysis for a $W' \to WH$ signal with $M_{W'} = 2$ TeV, including all $W$ and $H$ decays, is 0.061 for the $WH$ or $ZH$ selections, a bit smaller than the efficiency of 0.10 reported by the ATLAS Collaboration.

\begin{figure}[t]
\begin{center}
\begin{tabular}{ccc}
\includegraphics[height=5.2cm,clip=]{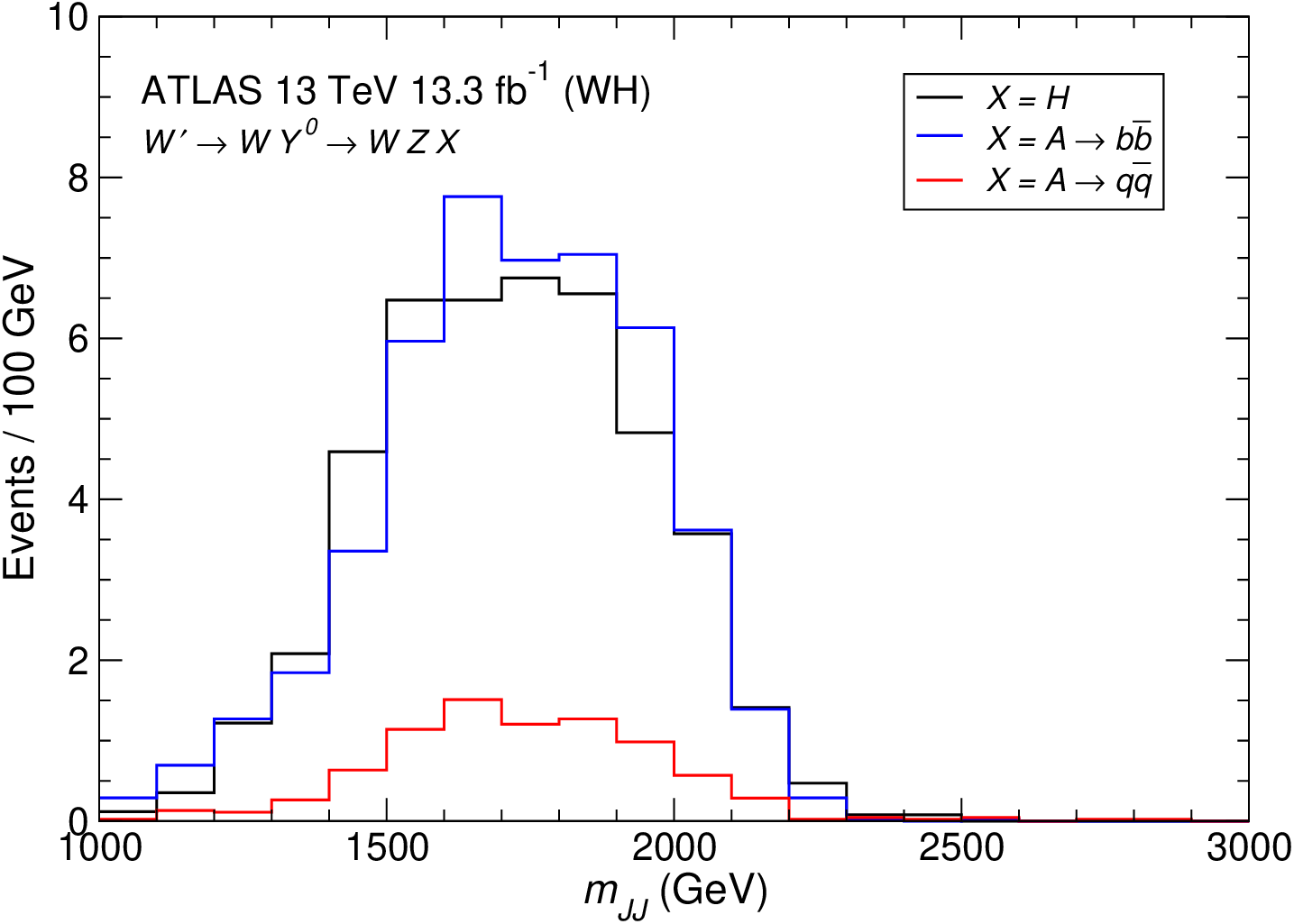} & \quad &
\includegraphics[height=5.2cm,clip=]{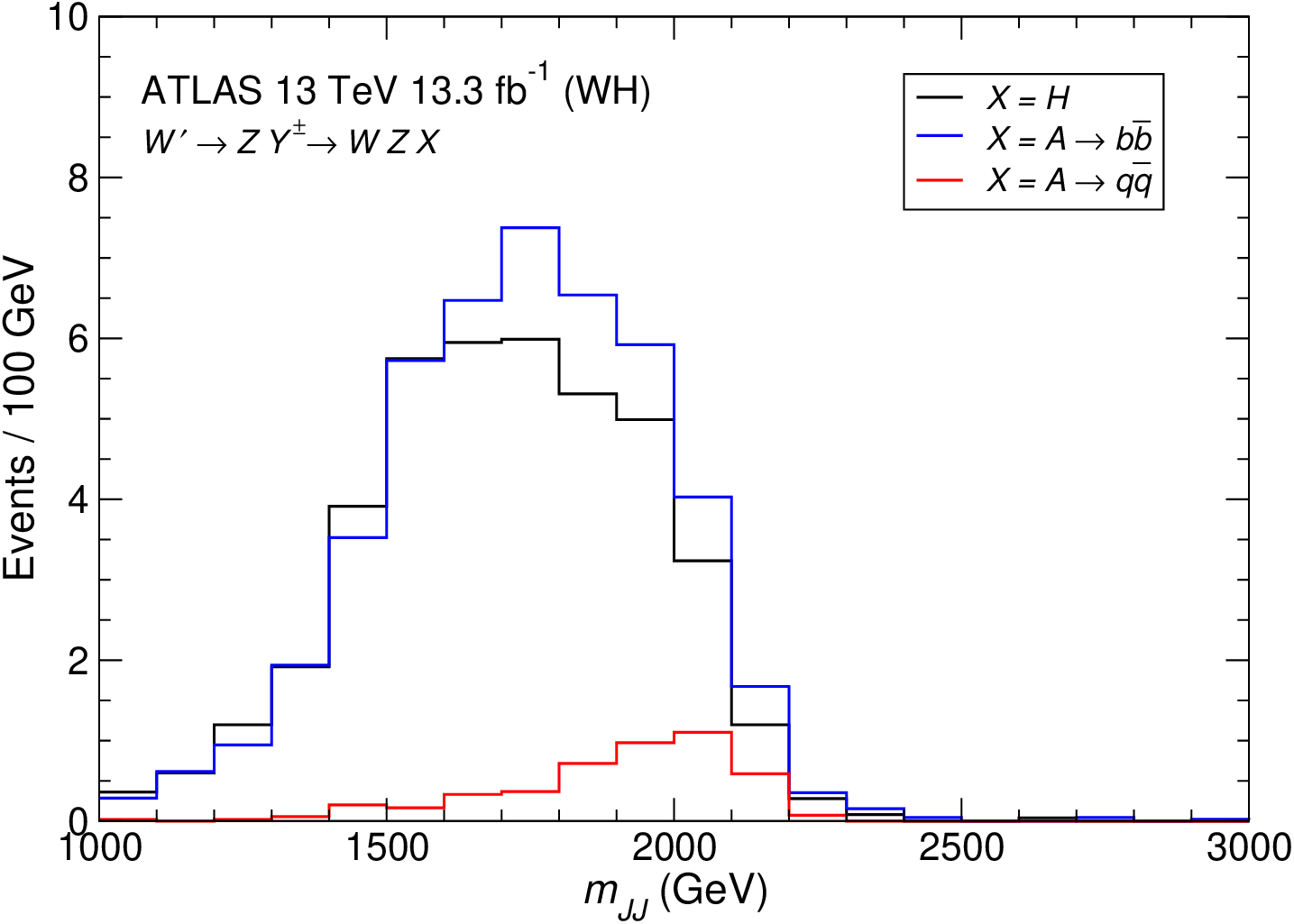} \\
\includegraphics[height=5.2cm,clip=]{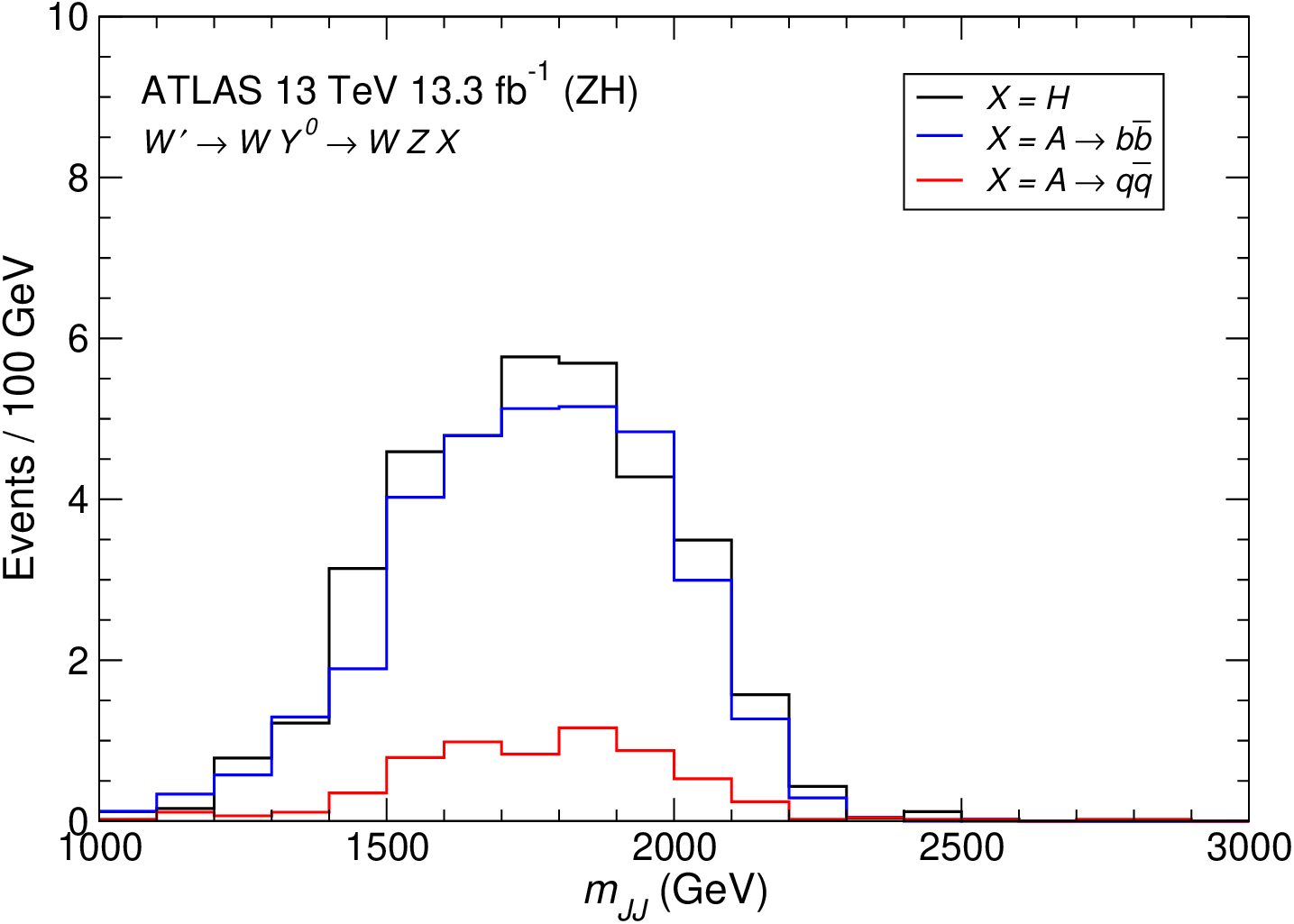} & \quad &
\includegraphics[height=5.2cm,clip=]{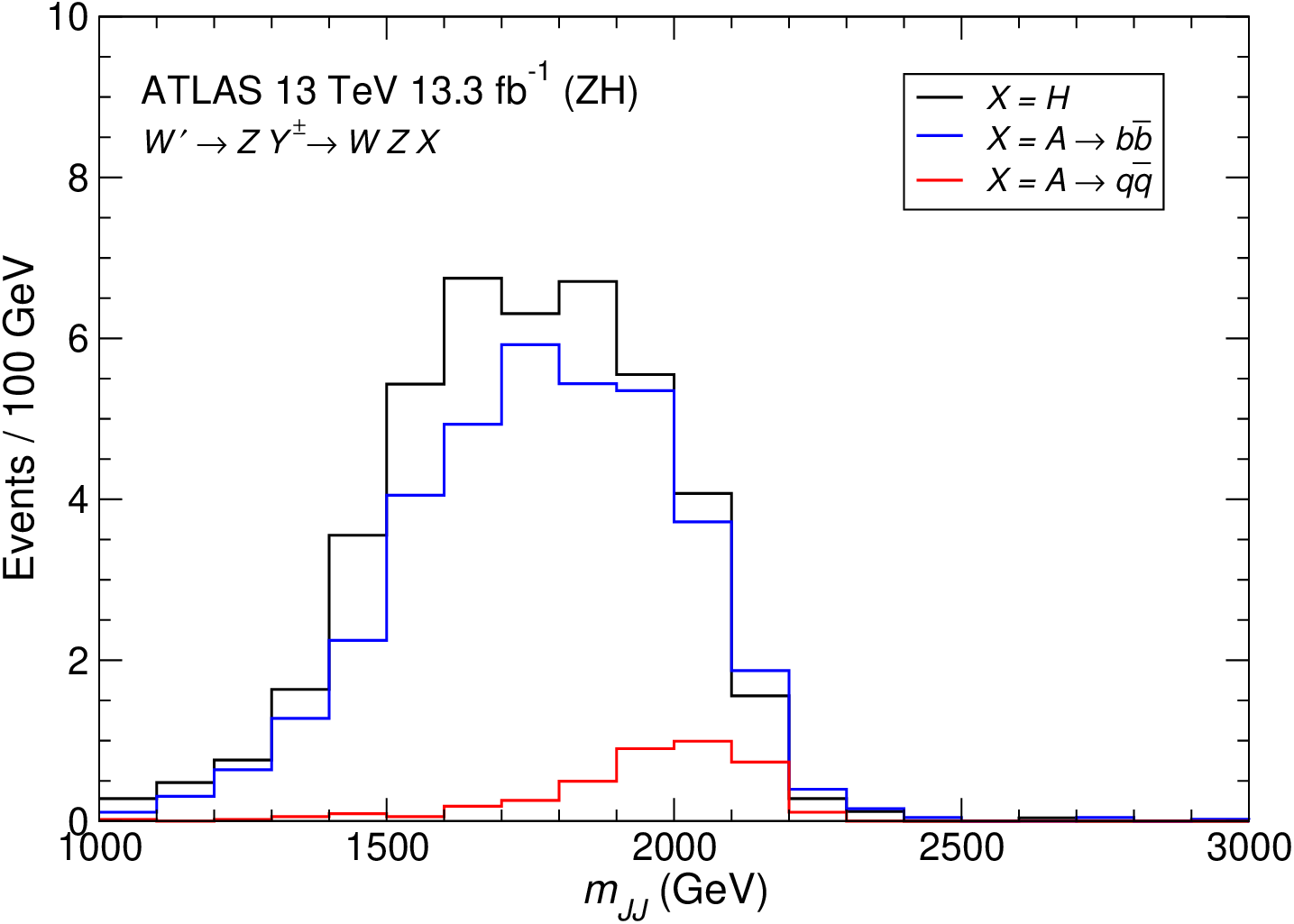}
\end{tabular}
\caption{Reconstructed invariant masses for the ATLAS hadronic $VH$ analysis~\cite{ATLAS:2016kxc}. Top: $WH$ selection with two $b$ tags. Bottom: $ZH$ selection with two $b$ tags. }
\label{fig:VHhad}
\end{center}
\end{figure}

We restrict our analysis to final states with two $b$ tags, which have the highest sensitivity, and present our results in figure~\ref{fig:VHhad}, with the same signal normalisation used throughout the paper but for a luminosity of 13.3 fb$^{-1}$. For $X = H$ and $X = A \to b \bar b$, in either $W'$ decay channel, the signals have about the right size and shape to explain the excesses found by the ATLAS Collaboration below 2 TeV, which reach $2.6\sigma$ in the $WH$ channel at $m_{JJ} = 1.6$ TeV.
In this regard, we point out that there is an apparent underfluctuation in the signal regions, compared to the data-driven prediction, for invariant masses around 1.8 TeV, especially in the two-tag $WH$ sample. Bearing in mind this effect, the predictions of the triboson scenarios with $X = H$ and $X = A \to b \bar b$ seem compatible with the measurements of the ATLAS Collaboration.

\end{document}